\documentclass[journal,10pt]{IEEEtran}
\title{Diagnosis/Prognosis of COVID-19 Images: Challenges, Opportunities, and Applications
\thanks{This Project was partially supported by the Department of National Defence's Innovation for Defence Excellence and Security (IDEaS)
program, Canada. Corresponding Author is Arash Mohammadi, email: arash.mohammadi@concordia.ca}}
\author{Arash Mohammadi$^{\dag}$,~\IEEEmembership{Senior Member,~IEEE}, Yingxu Wang$^{**}$,~\IEEEmembership{Fellow,~IEEE}, Nastaran Enshaei$^{\dag}$,~\IEEEmembership{Graduate Student Member,~IEEE}, Parnian Afshar$^{\dag}$,~\IEEEmembership{Graduate Student Member,~IEEE}, Farnoosh Naderkhani$^{\dag}$,~\IEEEmembership{Member,~IEEE},  Anastasia Oikonomou (MD)$^{\dag\dag}$, Moezedin Javad Rafiee (MD)$^{*}$, Helder C. R. Oliveira$^{**}$,~\IEEEmembership{Member,~IEEE}, Svetlana Yanushkevich$^{**}$,~\IEEEmembership{Senior Member,~IEEE}, and  Konstantinos N. Plataniotis$^{\ddag\ddag}$,~\IEEEmembership{Fellow,~IEEE}\\\vspace{.2in}
$^{\dag}$Concordia Institute for Information Systems Engineering (CIISE),  Concordia University, Montreal,~Canada \\\vspace{.05in}
$^{\dag\dag}$Department of Medical Imaging, Sunnybrook Health Sciences Centre, University of Toronto, Toronto,~Canada\\\vspace{.05in}
$^{*}$Department of Medicine and Diagnostic Radiology, McGill University Health Center-Research Institute \\\vspace{.05in}
$^{\ddag\ddag}$ Department of Electrical and Computer Engineering, University of Toronto, Toronto, Canada\\\vspace{.05in}
$^{**}$ Department of Electrical and Computer Engineering, University of Calgary, Calgary, AB, Canada
}
\usepackage{flushend}
\usepackage[dvipdfmx]{graphicx}
\usepackage{epsfig}
\usepackage{amsmath}
\usepackage{multirow}
\usepackage{amsthm}
\usepackage{amssymb }
\usepackage{amsmath}
\usepackage{bm}
\usepackage{cite}
\usepackage{float}
\usepackage{mathrsfs}
\usepackage{amsfonts}
\usepackage{algorithm, algpseudocode}
\usepackage{subfigure}
\usepackage[usenames]{color}
\usepackage[dvipsnames]{xcolor}
\usepackage{tcolorbox}
\usepackage{lipsum}
\usepackage{adjustbox}
\usepackage{lscape}
\usepackage{forest}
\usepackage{array}
\usepackage{lscape}
\usepackage{array}
\newcolumntype{P}[1]{>{\centering\arraybackslash}p{#1}}
\newcolumntype{M}[1]{>{\centering\arraybackslash}m{#1}}
\usepackage{pifont}% http://ctan.org/pkg/pifont
\usepackage{rotating}
\usepackage{booktabs}
\usepackage{longtable}
\usepackage{tablefootnote}
\usepackage{nccmath}

\usetikzlibrary{shadows}
\newcolumntype{C}[1]{>{\centering\arraybackslash}p{#1}}
\newcommand{\cmark}{\ding{51}}%
\begin{document}

\date{\today}
\maketitle
\thispagestyle{empty}
\vspace{-.2in}

%\tableofcontents
%OOOOOOOOOOOOOOOOOOOOOOOOOOOOOOOOOOOOOOOOOOOOOOOOOOOOOOOOO
\begin{abstract}
%OOOOOOOOOOOOOOOOOOOOOOOOOOOOOOOOOOOOOOOOOOOOOOOOOOOOOOOOO
The novel Coronavirus disease, COVID-19, has rapidly and abruptly changed the world as we knew in 2020. It becomes the most unprecedent challenge to analytic epidemiology in general and signal processing theories in specific. Given its high contingency nature and adverse effects across the world, it is important to develop efficient processing/learning models to overcome this pandemic and be prepared for potential future ones. In this regard, medical imaging plays an important role for the management of COVID-19. Human-centered interpretation of medical images is, however, tedious and can be subjective. This has resulted in a surge of interest to develop Radiomics models for analysis and interpretation of medical images. Signal Processing (SP) and Deep Learning (DL) models can assist in development of robust Radiomics solutions for diagnosis/prognosis, severity assessment, treatment response, and monitoring of COVID-19 patients. In this article, we aim to present an overview of the current state, challenges, and opportunities of developing SP/DL-empowered models for diagnosis (screening/monitoring) and prognosis (outcome prediction and severity assessment) of COVID-19 infection. More specifically, the article starts  by elaborating the latest development on the theoretical framework of analytic epidemiology and hypersignal processing for COVID-19. Afterwards, imaging modalities and Radiological characteristics of COVID-19 are  discussed. SL/DL-based Radiomic models specific to the analysis of COVID-19 infection are then described covering the following four domains: Segmentation of COVID-19 lesions; Predictive models for outcome prediction; Severity assessment, and; Diagnosis/classification models. Finally, open problems and opportunities are presented in detail. 
%OOOOOOOOOOOOOOOOOOOOOOOOOOOOOOOOOOOOOOOOOOOOOOOOOOOOOOOOO
\end{abstract}
%OOOOOOOOOOOOOOOOOOOOOOOOOOOOOOOOOOOOOOOOOOOOOOOOOOOOOOOOO

%OOOOOOOOOOOOOOOOOOOOOOOOOOOOOOOOOOOOOOOOOOOOOOOOOOOOOOOOO
\section{Introduction} \label{sec:Introduction}
%OOOOOOOOOOOOOOOOOOOOOOOOOOOOOOOOOOOOOOOOOOOOOOOOOOOOOOOOO
We are facing, first hand, an abrupt and sudden change of the world as we knew because of the novel Coronavirus outbreak. Known as COVID-19, it first appeared in late December 2019 in Wuhan, China, and few months later, on the 11$^{\text{th}}$ of March 2020, was characterized as a pandemic by World Health Organization (WHO). Given its high contingency nature, relatively unknown behaviour, systemic complications, and adverse effects ranging from human fatalities to economic recessions across the world, it is of importance to develop efficient processing/learning models to help overcome this pandemic and be prepared for potential future ones. The Reverse-Transcription Polymerase Chain Reaction (RT-PCR) is the standard testing approach for early diagnosis of suspected cases of COVID-19. Unavailability of enough RT-PCR testing kits particularly in areas severely affected by the pandemic and its relatively high and variable false-negative rate (i.e., highest during the first five days (up to 67\%), and lowest on day 8  (21\%)~\cite{Kucirka:2020}), resulted in focusing on medical image Radiomics~\cite{Afshar:2019} as a complementary source for diagnosis/prognosis.

Recent studies~\cite{Shireview, Dongreview, Jamshidireview}
show that chest Computed Tomography (CT) scans and Chest Radiographs (CXR) reveal informative features of COVID-19 that can assist in the monitoring, severity assessment, and treatment of COVID-19~\cite{Wang:2020-1}. According to the guideline provided by WHO, use of chest imaging as a complementary source of data is recommended in different scenarios and stages of COVID-19 to assist radiologists and physicians to detect and evaluate the disease more accurately. CT and CXR can decrease the false negative rate both at the admission and discharge times. It is worth mentioning that chest CT  has a key role for diagnosis of COVID-19 in the very early stages of the infection and also to set up a prognosis. Comparisons between CT and RT-PCR at early stages of COVID-19 infection show that  CT abnormalities may appear before PCR positivity. In other words, CT has greater sensitivity during the early stages of the infection. In addition, false negatives in RT-PCR results occur both at the admission and discharge.  Finally, CT plays its role over the course of the disease for evaluating changes in severity and for treatment adjustments. The key power of chest imaging is in its prognostic value to identify severity of the disease, likelihood of needing hospitalization and/or admission to Intensive Care Unit (ICU).  However, interpretation of chest images for confirming the suspected cases of COVID-19, and severity assessment of the disease based on imaging findings are  time-consuming and may be challenging.

%%%%%%%%%%%%%%%%%%%%%%%%%%%%%%%%%%%
\begin{figure}[t!]
\centering
\includegraphics[width=0.47\textwidth]{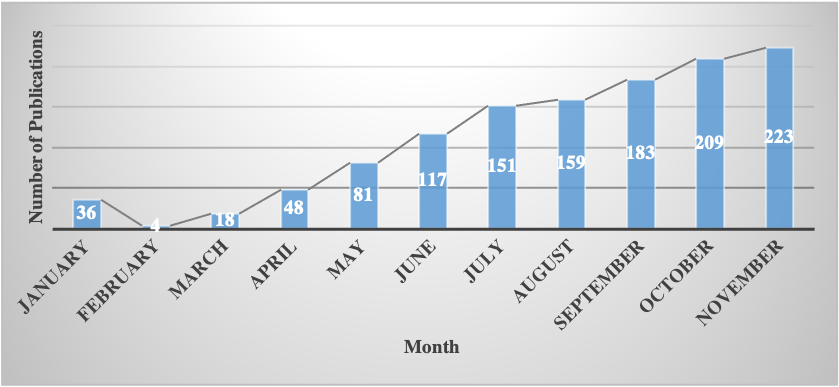}
\caption{\textcolor{black}{The trend in COVID-19-related research.}}\label{fig:trend}
\end{figure}
%%%%%%%%%%%%%%%%%%%%%%%%%%%%%%%%%%%
Interpretation of CT and CXR images should be performed by expert thoracic radiologists, who may not be easily accessible especially during the outbreak when the number of suspected cases of COVID-19 is growing up exponentially. To address these issues,  there has been a surge of interest in developing Signal Processing (SP) and Deep Learning (DL) techniques to extract informative features from chest images and help in fast detection and risk assessment of the COVID-19 infection. We would like to mention that although research works on COVID-19-related topics have started very recently, the extensive number of research works that have been disseminated during this short period of time makes the topic mature. More specifically, extensive research works on application of Signal/Image Processing and Artificial Intelligence (AI) for COVID-19 have lead to almost $1,200$ publications by the end of November 2020. Fig.~\ref{fig:trend} presents COVID-19 research trend in year 2020, obtained from PubMed with the keyword ``COVID-19'' and either of the following keywords: ``Signal Processing'', ``Machine Learning'', ``AI'', or  ``Deep Learning''. These publications cover several aspects and applications of SP/DL for COVID-19 including diagnosis, classification, detection, segmentation, severity assessment, and survival analysis. %This illustrates the maturity of COVID-19 topic and the extensive research works happening in this field.

In summary, in this feature article, we aim to present an overview of the current state, challenges, and opportunities of developing SP/DL-empowered models for diagnosis and prognosis of  the COVID-19 infection based on medical images. The article will mainly  focus on the problems, applications and on how SP/DL models can be used to address the identified problems/applications. In brief, we will cover the following main topics:
\begin{enumerate}
\item[(i)] We focus on SP techniques specific to  COVID-19 images and target specialized SP aspects of COVID-19 diagnoses, including \textit{Analytic Epidemiology}  and \textit{Hypersignal Processing (HP)} theory as advanced processing solutions of COVID-19.
\item[(ii)] Introduction of potential applications of SP/DL-based models for the diagnosis and predictive prognosis of COVID-19 infections using medical images as the main source of data.
\item[(iii)] Investigation of the required medical background related to COVID-19 for development of advanced SP/DL models. An overview of different characteristics of COVID-19 that can be observed on chest images is presented. Furthermore, we describe how these imaging findings are related to the severity of the disease, and how they can be utilized in the development of  SP/DL models.
\item[(iv)] Presentation of DL Radiomic directions specific to the analysis of COVID-19 infection. We focus on DL-based solutions from the following four different aspects: Segmentation of COVID-19 Lesions; Predictive models for outcome prediction in COVID-19 patients; DL/SP models for severity assessment, and; Diagnosis and classification of COVID-19 cases.
\item[(v)] Introducing challenges, open problems, and opportunities of developing intelligent and autonomous models for diagnosis/prognosis of COVID-19.
\end{enumerate}

\vspace{.1in}
\noindent
\textit{\textbf{Distinction with Existing Articles}}: As a final note, we would like to briefly elaborate on the differences between this article and a recent feature article~\cite{Afshar:2019} on Radiomics and other survays/tutorials~\cite{Shireview, Dongreview, Jamshidireview} on COVID-19. In particular, Reference~\cite{Afshar:2019} is focused on hand-crafted and deep learning-based techniques for extracting features from cancer-related images. Cancer diagnosis is essentially a completely different task than that of COVID-19 analysis, which requires its own specific techniques and solutions. For instance, while nodules have solid shapes and defined locations, COVID-19 infection areas could be multifocal, ill defined and diverse in pattern or (morphology). Therefore, the traditional Radiomics filters are not applicable to the latter. Furthermore, images from patients with pulmonary malignancies contain far less motion artifacts compared to COVID-19 ones, where patients suffer from dyspnea, calling for more advanced artifact reduction techniques. Deep learning models developed for Cancer Radiomics  are also not transferable to COVID-19 without modifications. This is mainly due to the fact that in a patient with COVID-19 a large number of slices may be affected, for which 3D analysis and more powerful resources are required. With regards to differences with recent survay/tutorial articles on COVID-19 research, Reference~\cite{Shireview} is limited to deep learning techniques. In this article, however, we also focus on SP modeling, applications, required medical background, and challenges/open-problems.  Reference~\cite{Dongreview} is more focused on the medical background and imaging modalities and AI models are not discussed in a separate section compared to this article, where different models and applications are separated and discussed. Reference~\cite{Jamshidireview} reviews a subset of deep learning models, without referring to SP methods, and challenges.

The reminder of the manuscript is organized as follows: Section~\ref{sec:HC_radiomics} is devoted to analytic epidemiology and hypersignal processing for COVID-19. Applications and imaging modalities for diagnosis/prognosis of COVID-19 images are then presented in Section~\ref{sec:Applicatios}. SL/DL-based Radiomic models specific to the analysis of COVID-19 infection are then described in Section~\ref{sec:DR} covering the following four application domains: Segmentation of COVID-19 lesions; Predictive models for outcome prediction; Severity assessment, and; Diagnosis/classification models. Finally, challenges, open problems, and opportunities are discussed in Section~\ref{sec:COO}. 
%OOOOOOOOOOOOOOOOOOOOOOOOOOOOOOOOOOOOOOOOOOOOOOOOOOOOOOO
\section{Analytic Epidemiology and HyperSignal Processing for COVID-19} \label{sec:HC_radiomics}
%OOOOOOOOOOOOOOOOOOOOOOOOOOOOOOOOOOOOOOOOOOOOOOOOOOOOOOO
In this section, we focus on SP techniques specific to  COVID-19 images (in Section~\ref{sec:DR}, we focus on DL models). Here, we target specialized SP aspects of COVID-19 diagnoses, including analytic epidemiology and Hypersignal Processing (HP) theory as advanced processing solutions of COVID-19. The worldwide outbreaks of COVID-19 and other contemporary contagious diseases have triggered a wide scope of transdisciplinary studies on epidemiology for their systematical treatments, control, prediction, prevention, management, and decision optimization~\cite{WHOGeneva,GoveCanada,CasesData}. The transdisciplinary investigations into the COVID-19 pandemic have led to the emergence of analytic epidemiology underpinned not only by epidemiology, biology, and medical sciences, but also by computer, big data, information, signal and sensor, AI, system science as well as mathematics, sociology, and economics.

%===================================================
\subsection{Analytic Epidemiology Models of COVID-19}
%===================================================
\noindent
\textit{Analytic epidemiology}~\cite{AndersonPopulation, Wang:2020-1} is a transdisciplinary study on the cognitive, theoretical, and mathematical models of COVID-19 and other contagious diseases. It is recognized that analytic epidemiology may be better studied by signal  explorations at the macro level rather than merely biological analyses at the micro level in order to not lose the forest for the trees~\cite{Wang:2020-1}.

The decision model of COVID-19 diagnoses may be formally described by a Cartesian product of the sets of symptoms~\cite{Wang:2020-1} and test results. Let the set of symptoms of COVID-19 be
\begin{equation}\medmath{
S \!=\! \Big\{\!S_1(Fever), S_2(Cough), S_3(BreathDifficulty), S_4(Chills), \nonumber}\vspace{-.1in}
\end{equation}
\begin{equation}\medmath{
~~S_5(ChillShaking),  S_6(MusclePain), S_7(HeadAche),\nonumber}
\vspace{-.1in}
\end{equation}
\begin{equation}\medmath{
\!\!\!\!\!\!\!\!\!\!\!\!\!\!\!\!\! S_8(SoreThroat), S_9(LossOfTaste/Smell)\Big\},\nonumber}
\end{equation}
and the set of lab tests be
\begin{equation}\medmath{
L = \Big\{L_1(NucleicAcid), L_2(SoreSample),
 L_3(LungImage) \Big\}.\nonumber}
\end{equation}
The diagnosis outcomes $E$ of COVID-19 infections are detected by the Cartesian product between the sets of logical values of \textit{detection symptoms}  $E_{S}$ and \textit{lab confirmations} $E_{L}$ as follows~\cite{Wang:2020-1}:
\begin{equation} \label{eq:W1} \medmath{
\begin{array}{l} {E\buildrel\wedge\over= E_{S} \times E_{L} =\mathop{\mbox{\large $\bm{R}$}{\rm \; }}\limits_{i=1}^{9} S_{i} {\rm |L}\times \mathop{\mbox{\large $\bm{R}$}{\rm \; }}\limits_{j=1}^{3} L_{j} {\rm |L}} \\ {{\rm \; \; \; }
\!\!\!\!\!\!\!\!\!\!\!\!\!\!\!=\left\{\begin{array}{l} {\!\!\![(\mathop{\wedge {\rm \; }}\limits_{i=1}^{9} S_{i} {\rm |L)}=T{\rm |L]}\wedge [(\mathop{\wedge {\rm \; }}\limits_{j=1}^{3} L_{j} {\rm |L)}=T{\rm |L]\; }{\rm //\; Positive}} \\
{\!\!\![(\mathop{\vee {\rm \; }}\limits_{i=1}^{9} S_{i} {\rm |L)}=T{\rm |L]}\wedge [(\mathop{\vee {\rm \; }}\limits_{j=1}^{3} L_{j} {\rm |L)}=T{\rm |L]\; //\; Suscetibly\; positive}} \\
{\!\!\![(\mathop{\vee {\rm \; }}\limits_{i=1}^{9} S_{i} {\rm |L)}=T{\rm |L]}\wedge [(\mathop{\wedge {\rm \; }}\limits_{j=1}^{3} L_{j} {\rm |L)}=F{\rm |L]\; }{\rm //\; Suscetibly\; negative}} \\
{\!\!\![(\mathop{\wedge {\rm \; }}\limits_{i=1}^{9} S_{i} {\rm |L)}=F{\rm |L]}\wedge [(\mathop{\wedge {\rm \; }}\limits_{j=1}^{3} L_{j} {\rm |L)}=F{\rm |L]\; }{\rm //\; Negative}} \end{array}\right. } \end{array} }
\end{equation}
where $T|L$ and $F|L$ denotes a Boolean logical variable for True or False, respectively. The diagnosing results are classified in the categories of symptomatic positive, susceptibly positive, negative, and susceptibly negative.
The big-R notation \cite{Wang2009, WangZatarain} is a generic calculus that denotes an iterative or recursive series of recurrent structures or embedded functions. Eq.~\eqref{eq:W1} reveals that many important symptoms and diagnosis of COVID-19 are in the domain of advanced signal processing as a foundation for COVID-19 diagnosis. In analytic epidemiology, the reproductive ratio $R_0$ of a contagious disease is modeled as an exponential transmission series $N_{inf}(t)$ on the $t_0 +$ \textit{k}th day, which is estimated by a product of initial infectives $N_{inf}(t)$ and the average reproductive rate raised to the \textit{k}th power:
\begin{equation} \label{eq:W2} \medmath{
N_{inf} {\rm (}t_{0} +k{\rm )}\buildrel\wedge\over= \bar{R}_{0} {}^{k} N_{inf} (t_{0} ),{\rm \; }\bar{R}_{0} >1.0,k\ge 0,N_{inf} (t_{0} )\ne 0 }.
\end{equation}
Therefore, the average reproductive rate of a pandemic transmission is reduced to the \textit{k}th root of the average ratio between the number of infectives $N_{in\!f}(t_{0}+k)$ cumulatively infected at $t_{0} + k$ by each initial infective $N_{in\!f}(t_0)$:
\begin{equation} \label{eq:W3}
\; \bar{R}_{0} \buildrel\wedge\over= {}^{k} \sqrt{\frac{N_{in\!f} {\rm (}t_{0} +k{\rm )}}{N_{in\!f} (t_{0} )} } ,{\rm \; }k\ge 0,N_{in\!f} (t_{0} )\ne 0
\end{equation}
For instance, WHO has empirically estimated $\bar{R}_{0}$ of COVID-19 in the range of $2.24$ to $4.00$~\cite{WHOGeneva}, which was considerably higher than those obtained in rigorous analyses with real-world signals according to Eqs.~\eqref{eq:W2} and~\eqref{eq:W3} as rigorously treated a long series of pandemic signals.

The \textit{reproductive rate} $\bar{R}_{0} $ in analytic epidemiology has been adopted as the key indicator \textit{$\theta$} for the congruous severity classified in two categories by the threshold $\bar{R}_{0} $= 1.0, i.e.:
\begin{equation} \label{eq:W4}
\theta =\left\{\begin{array}{l} {congruous,{\rm \; \; \; \; \; }R_{0} (t)>1.0} \\ {incongruous,{\rm \; \; 1.0}\ge R_{0} (t)\ge 0} \end{array}\right.
\end{equation}
However, when investigating into the nature of pandemic dynamics to rigorously predict the pandemic trends, we found that in order to model more general and complex pandemic dynamics, the reproductive rate must be treated as a series of variables \textit{R${}_{0}$}(\textit{t}) over time. This finding has led to the formal model of the series of dynamic\textit{ reproductive rates }of COVID-19, which is recursively determined by a long-chain of causal probabilities over time:
\begin{equation} \label{eq:W5} \medmath{
\mathop{\mbox{\large $\bm{R}$}}\limits_{t=2}^{n} {\rm \; }R_{0} (t-1)\buildrel\wedge\over= \mathop{\mbox{\large $\bm{R}$}}\limits_{t=2}^{n} {\rm \; }\frac{N_{in\!f} {\rm (}t-1{\rm )}}{N_{in\!f} (t-2)} ,{\rm \; }R_{0} (0)=1,N_{in\!f} (t)\ne 0 }.
\end{equation}
Simulations performed based on real-world signals have provided highly accurate predications based on the mathematical model of the analytic epidemiology theory and its dynamic predictability for the pandemic signal series~\cite{Wang:2020-1}.

%===================================================
\subsection{Hypersignal Processing (HP) for COVID-19 Image Diagnoses}
%===================================================
Hypersignals are a general structure of abstract or real-world signals beyond 1D or its parallel compositions~\cite{GoveCanada,CasesData}. Hypersignal Processing (HP) theory provides a unified mathematical model for advancing 1D signal (voice and time series) processing to 2D (images) and nD (generic hypersignals) processing~\cite{Wang2009, WangZatarain, WangKeynoteNeuro, WangKeynoteGeneration, WangKeynoteCognitive}. The hypersignals may be embodied by sequences of images (videos), language expressions and semantics, knowledge structures, neural networks, and AI systems. Therefore, HP demands novel theories, mathematical means, and algorithms~\cite{WangKeynoteNeuro, WangKeynoteGeneration, WangKeynoteCognitive}.

For instances, the HP theory models hypersignals as follows:
\begin{equation} \label{eq:W6}
\left\{\begin{array}{l} {Text{\rm |TX:\; TX}\buildrel\wedge\over= S} \\ {Voice{\rm |V:\; V}\buildrel\wedge\over= B\times T} \\ {Image{\rm |I:\; I}\buildrel\wedge\over= B\times B} \\ {Video{\rm |M:\; M}\buildrel\wedge\over= B\times B\times T} \\ {HyperSig{\rm |H:\; H}\buildrel\wedge\over= \mathop{\mbox{\large $\bm{R}$}}\limits_{i=0}^{n_{i} } {\rm \; }\mathop{\mbox{\large $\bm{R}$}}\limits_{j=0}^{n_{j} } {\rm \; \ldots\; }\mathop{\mbox{\large $\bm{R}$}}\limits_{k=0}^{n_{k} } {\rm \; }\Theta (i,j, \ldots, k)} \end{array}\right.
\end{equation}
where \textit{B} stands for a byte, \textit{T} for time, \textit{S} for a string, and $\Theta$ for an instance of an abstract hypersignal.

More generically, the hyper Structure Model (SM) is introduced to model complex hyper signals and entities. For example, the SM model of the color scheme of an image signal is formally modeled as:
\begin{eqnarray} \label{eq:W7}
Image |\text{SM} \buildrel\wedge\over=
\left\{\begin{array}{l}
Image|\text{FG}=\mathop{\mbox{\large $\bm{R}$}}\limits_{i=1}^{|X|} \mathop{\mbox{\large $\bm{R}$}}\limits_{j=1}^{|Y|} Pixel (i,j)|\text{PG} \\
Image |\text{FB}=\mathop{\mbox{\large $\bm{R}$}}\limits_{i=1}^{|X|} \mathop{\mbox{\large $\bm{R}$}}\limits_{j=1}^{|Y|} Pixel (i,j)|\text{PB} \\
Image |\text{FR}^* = \mathop{\mbox{\large $\bm{R}$}}\limits_{i=1}^{|X|} \mathop{\mbox{\large $\bm{R}$}}\limits_{j=1}^{|Y|} Pixel (i,j) |\text{PR}^*\\
Image |\text{FG}^*=\mathop{\mbox{\large $\bm{R}$}}\limits_{i=1}^{|X|} \mathop{\mbox{\large $\bm{R}$}}\limits_{j=1}^{|Y|} Pixel (i,j) |\text{PG}^* \\
Image |\text{FB}^* = \mathop{\mbox{\large $\bm{R}$}}\limits_{i=1}^{|X|} \mathop{\mbox{\large $\bm{R}$}}\limits_{j=1}^{|Y|} Pixel (i,j) |\text{PB}^*
\end{array}\right.  \!\!\!\!\!\!\!\!\!\!\!\!\!\!\!\! \nonumber\\
\end{eqnarray}
where the 2D frame is represented in six forms including FC (color), FG (gray), FB (black/white), FR* (red), FG* (green), FB* (blue), and the composition FC = (FR*, FG*, FB*).

A paradigm of HP towards COVID-19 is represented by Image Frame Algebra (IFA)~\cite{Wang2009, WangKeynoteCognitive}, which processes and diagnoses suspectedly infected lung images by efficient and accurate hypersignal handling according to a set of IFA operators. In IFA, a generic image model is an SM based on Eq.~\eqref{eq:W7}. According to IFA, the differential algorithm for COVID-19 images manipulations is implemented as follows:
\begin{eqnarray} \label{eq:W8}
\delta (I_{1},I_{2}) & \buildrel\wedge\over=&1-\sigma (I_{1}^{} ,I_{2}^{} ),{\rm \; }p_{1,2} (i,j) |\text{PG}\in [0, 255] \nonumber\\
&=&\frac{\sum _{i=1}^{|X|}\sum _{j=1}^{|Y|}\left|p_{1} (i,j)|\text{PG} - p_{2} (i,j) |\text{PG}\right|}{255|X|\bullet |Y|} \nonumber\\
\end{eqnarray}
The operations of image differentiation in IFA may be expressed according to Eq.~\eqref{eq:W8} as follows:
\begin{eqnarray} \label{eq:W9}
\left\{\begin{array}{ll}
\text{Space Differentiation:} & Dif\!f_{s} \buildrel\wedge\over= \delta_{s} (I_{L}, I_{R}) \\
\text{Time Differentiation:} & Dif\!f_{t} \buildrel\wedge\over= \delta_{t} (I_{t_{1} }, I_{t_{1} })
\end{array}\right.
\end{eqnarray}
where the first model expresses a differentiation between the left and right images of a symmetric structure such as lungs, breasts, and the brain. The second model denotes a sequential differentiation of image series with respect to time.

\begin{figure}[t!]
\centering
\includegraphics[scale = .7]{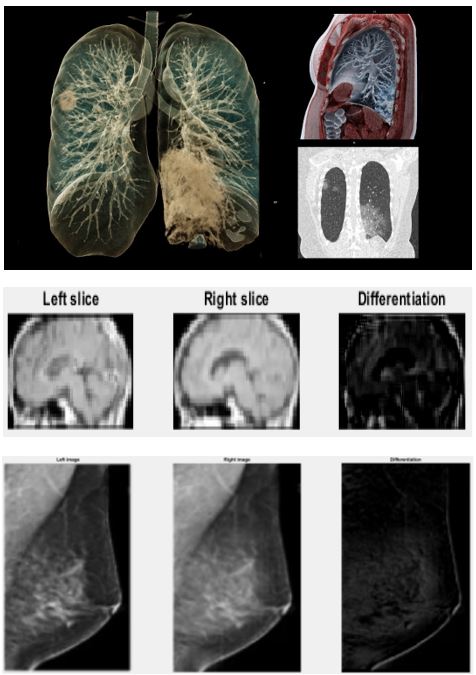}
\caption{Formal diagnoses of COVID-19 affected lung images and other applications
by image differentiations.}
\label{fig:Wim-dif}
\vspace{-.1in}
\end{figure}

For instances, the results of COVID-19 affected lung images, brain tumors, and breast tumors may be diagnosed according to the generic image differential algorithm (Eqs.~\eqref{eq:W8} and~\eqref{eq:W9}), respectively, as shown in Fig.~\ref{fig:Wim-dif}.

%%%%%%%%%%%%%%%%%%%%%%%%%%%%%%%%%%%%%%
%%%%%%%%%%%%%%%%%%%%%%%%%%%%%%%%%%%%%%

%OOOOOOOOOOOOOOOOOOOOOOOOOOOOOOOOOOOOOOOOOOOOOOOOOOOOOOO
\section{Applications and Imaging Modalities for Diagnosis/Prognosis of COVID-19 Images} \label{sec:Applicatios}
%OOOOOOOOOOOOOOOOOOOOOOOOOOOOOOOOOOOOOOOOOOOOOOOOOOOOOOO
As stated previously, chest imaging provides an important source of data for diagnosis/prognosis of COVID-19 infection, assessment of treatment response, and monitoring of  COVID-19 patients.  In brief, several recent studies have been conducted to investigate specific characteristics of Coronavirus disease on chest images that can be used for design of processing/learning models. Different types of chest imaging patterns and distribution of lung involvement are related to the severity/stage of the COVID-19 infection and can help construct predictive SP/DL models to make decisions on hospital admission versus home isolation, non-ICU hospital admission versus ICU admission, on monitoring the treatment process and on the time of home discharge.

In this section, we first present an overview of potential applications of SP/DL models for the diagnosis and predictive prognosis of COVID-19 infections. Then, we present  the required medical background related to COVID-19 for development of SP/DL models and will review different imaging modalities. Advantages and limitations of each modality is discussed together with its application in different stages of COVID-19 management. Generally speaking, applications of SP/DL models for COVID-19 diagnosis/prognosis can be classified into the following four categories:

\vspace{.1in}
\noindent
\textit{\textbf{$\bullet$ Diagnosis of COVID-19 Pneumonia from other Community Acquired Pneumonia (CAP):}} The most common COVID-19 symptoms (cough,  shortness of breath, and fever) overlap with CAP symptoms. CAP is mainly caused by a bacterial infection but can also be caused by viruses. In most cases, microbiological tests for CAP such as cultures of sputum and blood, are time consuming with poor sensitivity and specificity, and are not enough to identify the main pathogen~\cite{metlay2020treatment}.
Thus, the Polymerase Chain Reaction (PCR) test via nasopharyngeal or oropharyngeal swab has been used for correct identification of the source of the viral cause of CAP (like influenza)~\cite{burk2016viral}.
For COVID-19, a variation of PCR test, the RT-PCR, has been used as a gold standard~\cite{abduljalil2020laboratory}.
Due to the test inaccuracy, the decision on treatment may be incorrect, i.e., antibacterial drug therapy can be administrated to all CAP patients who may not be confirmed COVID-19 positive, while for patients tested positive for COVID-19 this treatment is not required~\cite{metlay2020treatment}. Cases of negative RT-PCR with persistent COVID-19 symptoms are submitted to chest imaging evaluation. This application domain will be further discussed in Sub-section~\ref{subsec:Calss}, where different SP/DL models developed for COVID-19 diagnosis are presented.

\vspace{.1in}
\noindent
\textit{\textbf{$\bullet$ Localizing COVID-19 Lesions and Identifying their Types:}} The pattern and extent of chest imaging findings is related to the stage and severity of  COVID-19 and  affects the treatment decision making. In Sub-section~\ref{subsec:Seg}, we will further describe applications of SP/DL models in localizing involved areas and demonstrating imaging features on CT scans.

\vspace{.1in}
\noindent
\textit{\textbf{$\bullet$ Outcome Prediction (COVID-19 Prognosis):}} To efficiently manage the limited medical resources during the pandemic, it is vital to accurately predict the risk of poor outcomes in COVID-19 patients. Some essential outcomes in COVID-19 patients are as follows:
\begin{itemize}
\item Mortality risk;
\item Progression to severe/critical stage;
\item Need for ICU admission/mechanical ventilation, and;
\item Length of hospital stay.
\end{itemize}
Predictive models are required to compute the probability of poor outcomes to help health-care professionals deliver appropriate services to high-risk patients. This application domain will be presented in detail in Sub-section~\ref{subsec:OP}.

\vspace{.1in}
\noindent
\textit{\textbf{$\bullet$ Severity Assessment of COVID-19:}} Chest imaging can be used to assess the lung infection severity in COVID-19 patients~\cite{metlay2020treatment}.   Calculation of percentage of parenchymal involvement and CT severity score can be achieved by segmenting the infected regions and lung areas in chest images. This  is required to evaluate and quantify severity, then prediction of prognosis of the COVID-19 infection. In Sub-section~\ref{subsec:SA}, we will present different SP/DL models developed for computing lung infection rate and CT severity score metrics as two commonly used criteria for severity assessment of COVID-19 infection.

%====================================================
\subsection{Imaging Modalities and Radiological Characteristics of COVID-19}
%====================================================
The Fleischner Society and the American College of Radiology, among others, recommend CT scan and CXR for COVID-19 patients with moderate to severe cases~\cite{rubin2020role}. Among the chest imaging modalities, the CXR is less sensitive, and less specific compared to CT. The advantages of CXR over CT include its fast availability, ease of execution, and minimization of in-hospital transmissions. In addition, the CXR findings correlate well with CT findings~\cite{ko2020pulmonary}. In some situations where a fast assessment is necessary, the Point-of-Care lung Ultrasound (POCUS) offers a radiation-free imaging modality with higher accuracy in patients without any previous cardiopulmonary disease~\cite{haak2020diagnostic}. In what follows, application of the above-mentioned imaging modalities for COVID-19 diagnosis/prognosis will be discussed in detail.

%--------------------------------------------------------------------------------------------------------------
\subsubsection{Computerized Tomography (CT) Scan}
\begin{figure}[t!]
\centering
\includegraphics[scale=0.5]{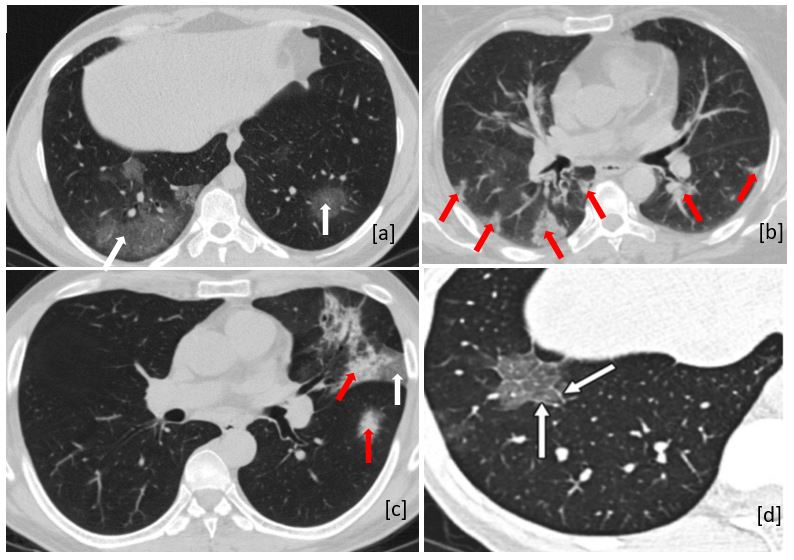}
\caption{The most common CT patterns in COVID-19 patients. (a) Axial CT image of a 38 year-old man with bi-lateral GGOs distributed in posterior lung regions~\cite{COVID-DATA}. (b) Axial CT image of a 60-year-old woman, scattered consolidation patterns with mainly peripheral distribution~\cite{COVID-DATA}. (c) Axial CT image of a 51-year-old man, the appearance of GGOs (white arrow) and consolidations (red arrows)~\cite{COVID-DATA}. (d) Crazy-paving pattern in axial CT image of a 43-year-old woman~\cite{bernheim2020chest}. All CT images have been obtained without contrast enhancement.}
    \label{fig:CTpatterns}
\end{figure}

%--------------------------------------------------------------------------------------------------------------
There has been considerable attention on CT imaging as the most useful imaging modality for representing COVID-19 infections. A study~\cite{li2020coronavirus} on $51$ patients with positive nucleic acid testing reported that only $3.9$\% of patients were misdiagnosed based on their chest CT images.
Fig.~\ref{fig:CTpatterns} shows common CT patterns in COVID-19 patients, where the most prevalent are ``\textit{Ground Glass Opacities (GGOs)}'' and ``\textit{Consolidations}''. GGO is a hazy transparent opacity that does not conceal lung vessels and bronchial areas~\cite{hansell2008fleischner}.
In a consolidation pattern, the air in the alveoli and peripheral bronchioles is replaced by fluid such as pus, water, blood, or an inflammatory material, obscuring the underlying distal airways and vascular margins~\cite{hansell2008fleischner}.
In a research study on $645$ confirmed COVID-19 patients, $88$\% of patients showed either pure GGOs or consolidation or both~\cite{zhang2020epidemiological}.
The appearance of pure GGO is more common in the early stage of the disease, while the appearance of GGOs with consolidations is more frequently seen in the progressive stage~\cite{sun2020systematic}. Another common CT pattern associated with COVID-19 is the so-called ``\textit{Crazy Paving}'' referring to thickened interlobular septa and intralobular interstitium superimposed on GGOs~\cite{hansell2008fleischner}.
The crazy paving pattern is more commonly seen in the progressive stage of the disease~\cite{sun2020systematic}. The appearance of the crazy paving/consolidation patterns as a sign of disease progression/severity can help radiologists evaluate the disease stage. Interlobular septal thickening, air bronchogram, and vascular enlargement are other CT findings in COVID-19 patients~\cite{sun2020systematic}.

%%%%%%%%%%%%%%%%%%%%%%%%%%%%%%%%%%%%%%%
%%%%%%%%%%%%%%%%%%%%%%%%%%%%%%%%%%%%%%%
\vspace{.05in}
\noindent
\textbf{\textit{Distribution of Lung Involvement in COVID-19:}} CT findings of COVID-19 infections demonstrate that most of the COVID-19 patients have had ``\textit{Bilateral}''  and ``\textit{multifocal}'' lung involvements. Bilateral involvement means that the lesions are distributed in both the right and left lungs, and multifocal involvement implies that more than one lobe (from five lobes) of the lung is affected by the disease. A systematic review of COVID-19 imaging findings~\cite{sun2020systematic} declared that in $17$ out of $36$ studies ($78.2$\%), the number of patients with bilateral lung involvement had been higher than the patients with unilateral involvement.

Researches also showed that COVID-19 lesions in most of the cases are distributed in lower lobes and have a ``\textit{Peripheral}'' instead of central distribution~\cite{sun2020systematic, bernheim2020chest}. Similarities between CT features of COVID-19 with other viral pneumonia pose limitations in using CT images to diagnose COVID-19. However, in a study of $58$ patients~\cite{bai2020performance}, six of seven radiologists could distinguish COVID-19 from other types of viral infections with an accuracy of $67-93$\% and a specificity of $93-100$\%. Peripheral distribution and GGOs were the most critical characteristics for distinguishing COVID-19 from non-COVID-19 pneumonia~\cite{bai2020performance}.

\vspace{.05in}
\noindent
\textbf{\textit{Correlation between CT Findings and Severity/Stage of the Disease:}}
Since CT images provide high sensitivity for detecting COVID-19 patients, they are reliable for developing SP/DL-based diagnosis models. Detecting the pattern of pulmonary involvement in COVID-19 including consolidation and/or crazy-paving patterns in chest CT images by SP/DL-powered networks, can help evaluate the disease severity. Quantifying the extent of lung involvement in COVID-19 patients is a deterministic criterion for assessing the disease's stage/severity. Different CT severity measures have been introduced in the literature that can be mapped to disease severity. The manual calculation of these measures by radiologists is tedious and time-consuming. Severity measures that can be automatically quantified by SP/DL models are as follows:
%
%%%%%%%%%%%%%%%%%%%%%%%%%%%%%%%%%%%%%%%
\begin{table}[t!]
\centering
\caption{\label{tab:PO}\textcolor{black}{Correlation between Percentage of Opacity (PO) measure and COVID-19 stag~\cite{HuangHan}.}}
\renewcommand{\arraystretch}{2}
\begin{adjustbox}{width=0.5\textwidth}
\begin{tabular}{|c|c|}
\hline
\textbf{COVID-19 stage} & \textbf{Percentage of Opacity (PO)} \\
\hline
Moderate & 2.2 (0.4, 7.1) (median and interquartile range).\\
\hline
Sever & 28.9  $\pm$  19.2 (mean  $\pm$  std). \\
\hline
Critical & 49.6  $\pm$  14.8 (mean  $\pm$  std). \\
\hline
\end{tabular}
\end{adjustbox}
\end{table}
%%%%%%%%%%%%%%%%%%%%%%%%%%%%%%%%%%%%%%%
\begin{itemize}
\item \textbf{\textit{Percentage of Opacity (PO):}} The PO measures the volume of COVID-19 abnormalities related to the whole lung volume. It was reported in Reference~\cite{HuangHan}  that the POs for COVID-19 patients are divided into three different categories as shown in Table~\ref{tab:PO}.
\item \textbf{\textit{Percentage of High Opacity (PHO) and Lung High Opacity Score (LHOS):}} The PHO and LHOS introduced in~\cite{ChagantiGrenier} quantify the volume of consolidation regions in the whole lung and across the lobes, respectively.
%%%%%%%%%%%%%%%%%%%%%%%%%%%%%%%%%%%%%%%
\begin{table}[t!]
\centering
\caption{\label{tab:CTscore}\textcolor{black}{Scoring system for measuring CT severity score~\cite{francone2020chest}.}}
\renewcommand{\arraystretch}{2}
\begin{adjustbox}{width=0.4\textwidth}
\begin{tabular}{|c|c|}
\hline
\textbf{Lobe involvement rate} & \textbf{Score} \\
\hline
\multicolumn{1}{|c|}{No involvement} & 0\\
\multicolumn{1}{|c|}{Involvement of less than 5\%.} & 1 \\
\multicolumn{1}{|c|}{Involvement from 5\% to 25\%.} & 2 \\
\multicolumn{1}{|c|}{Involvement from 26\% to 50\%.} & 3 \\
\multicolumn{1}{|c|}{Involvement from 51\% to 75\%.} & 4 \\
\multicolumn{1}{|c|}{Involvement is higher than 75\%.} & 5 \\
\hline
\end{tabular}
\end{adjustbox}
\end{table}
%%%%%%%%%%%%%%%%%%%%%%%%%%%%%%%%%%%%%%%
\item \textbf{\textit{CT Severity Score:}} Authors in~\cite{francone2020chest} used a severity measure for COVID-19 patients, referred to as the CT score, that measures the extent of involvement based on a semi-quantitative scoring for each of the five lobes. The score ranges from $0$ to $5$, which is computed as shown in Table \ref{tab:CTscore}. The overall CT score would be between $0$ and $25$, which is the sum of the lobar scores (some studies use a different scoring scale, which ends up to a CT score between $0$ and $20$~\cite{bernheim2020chest}). Francone, \textit{et al.}~\cite{francone2020chest} conducted research on $130$ COVID-19 patients and evaluated the correlation between CT score and disease severity. They showed that the CT score is strongly correlated with the COVID-19 clinical stage and severity. For patients in severe or critical categories, CT score is significantly higher than patients in the mild category~\cite{francone2020chest}. CT score greater than $18$ (out of 25) can be used as a predictor of mortality in COVID-19 patients~\cite{francone2020chest}.  CT score is highly correlated with patients' age. In~\cite{francone2020chest}, authors revealed that CT score in patients with age range $>50$ was significantly higher than those in the age range of $26$-$50$. For patients in late-stages of the disease, the CT score is higher than those in early-stages. CT score, together with patients' age, can be used to predict COVID-19 patients' death.
\end{itemize}
%

%--------------------------------------------------------------------------------------------------------------
\subsubsection{Chest Radiography (CXR)}
%--------------------------------------------------------------------------------------------------------------
Some studies report that CXR images often show no lung infection in COVID-19 patients at early stages resulting in a low sensitivity of $69$\% for diagnosis of COVID-19~\cite{wong2020frequency}. However, CXR is helpful for prediction of clinical outcome and for detection of COVID-19 in areas with limited access to reliable RT-PCR testing kits. The most commonly observed patterns in CXR of COVID-19 patients are GGOs and consolidations with bilateral peripheral distribution~\cite{wong2020frequency}. Pre-existence of medical conditions such as heart or other lung diseases will make the interpretation of CXR images challenging. Therefore, the interpretation of CXRs in younger patients would be more reliable and predictive. In Reference~\cite{ToussieVoutsinas}, the authors developed a scoring approach for severity assessment and outcome prediction of COVID-19 patients between the ages of $21$ to $50$ years based on their CXR images. In their scoring system, each lung is divided into three zones. A binary score is then given to each zone based on the appearance/absence of COVID-19 abnormalities, and the total score would be in the range of $0$-$6$. Their study on $338$ patients demonstrates that there is a significant correlation between CXR score greater than two and hospital admission. They also reported that a CXR score greater than three could predict the need for intubation. Using lung Edema severity measure, referred to as RALE score, the authors in~\cite{cozzi2020chest} quantify the extent of lung involvement and compute correlations with the risk of ICU admission for COVID-19 patients. Recent research works have demonstrated potentials of developing SP/DL-based models for grading the disease stage and performing outcome-prediction using CXR images.

%--------------------------------------------------------------------------------------------------------------
\subsubsection{Ultrasound}
%------------------------------------------------------------------------------------------------
Beside the advantages of using CT or CXR combined with RT-PCR test for a correct and precise diagnosis of COVID-19, these imaging modalities have limitations, including diagnostic accuracy, logistic challenges, time-consuming assessment and the use of ionizing radiation~\cite{haak2020diagnostic}. Despite low sensitivity of Ultrasound for diagnosis of COVID-19 patients in mild and moderate categories, lung ultrasound has shown high-sensitivity results in critical cases~\cite{lu2020clinical}.
Due to its low cost, portability, ease of use, and being radiation-free, lung ultrasound can play a crucial role in the follow up and monitoring patients in the ICU. Furthermore, Ultrasound has been widely used for the diagnosis and monitoring of COVID-19 in pregnant women. In Italy, health professionals used lung ultrasound as a screening tool and developed a lung ultrasound score for evaluating the severity of the disease in COVID-19 patients~\cite{vetrugno2020our}.

In another study with $93$ patients, where $27$ ($29$\%) of them were tested positive for COVID-19 by RT-PCR or CT, the Ultrasound imaging achieved a sensitivity of 89\% and specificity of 59\% \cite{haak2020diagnostic}. Considering a subgroup of 37 patients without any cardiopulmonary disease, the assessment based on Ultrasound revealed and sensitivity of 100\% and specificity of 76\% \cite{haak2020diagnostic}. Thus, Ultrasound represents a valuable imaging modality for the detection or assessing COVID-19 severity mainly in patients without any medical history of cardiopulmonary disease.

%%%%%%%%%%%%%%%%%%%%%%%%%%%%%%%%%%%
\begin{table*}[t!]
\centering
\caption{\label{tab:datasets}\textcolor{black}{Available COVID-19 CT scan datasets.}}
\renewcommand{\arraystretch}{2}
\begin{adjustbox}{width=1\textwidth}
\begin{tabular}{c c|c|c|c|c|c|c|c|c|c|c|c|}
\cline{2-13}
 & \multicolumn{3}{|c|}{\textbf{Number of cases}}
 & \multicolumn{2}{|c|}{\textbf{Label type}}
 & \multicolumn{2}{|c|}{\textbf{Data Source}}
 & \multicolumn{2}{|c|}{\textbf{CT volume}}
  & \multicolumn{3}{|c|}{\textbf{Label Level }}
 \\
\hline
 \multicolumn{1}{|c|}{Dataset} & COVID & CAP & Normal  & Classification & Segmentation  & Multiple & Single  & Available & Not available & Patient-level & Slice-level & Lobe-level \\
\hline
\multicolumn{1}{|c|}{Reference~\cite{Bjorke:2020}} & 49 & NA \footnotemark & NA & &\cmark  & \cmark & & \cmark & & &\cmark & \\
\hline
\multicolumn{1}{|c|}{Reference~\cite{Jun:2020}} & 20 & NA & NA & &\cmark  & \cmark & & \cmark & & &\cmark & \\
\hline
\multicolumn{1}{|c|}{Reference~\cite{Cohen:2020}} & 20 & NA & NA & &\cmark  & \cmark & & \cmark & & &\cmark & \\
\hline
\multicolumn{1}{|c|}{Reference~\cite{Morozov:2020}} & 856 & NA & 254 & \cmark &  & \cmark & & \cmark & & \cmark& & \\
\hline
\multicolumn{1}{|c|}{Reference~\cite{Zhao:2020}} & 216 & NA & 55 & \cmark &  & \cmark & &  &\cmark & &\cmark & \\
\hline
\multicolumn{1}{|c|}{Reference~\cite{Soares:2020}} & 60 & NA & 60 & \cmark &  & \cmark & &  &\cmark & &\cmark & \\
\hline
\multicolumn{1}{|c|}{Reference~\cite{Rahimzadeh:2020}} & 95 & NA & 282 & \cmark &  &  &\cmark & \cmark & & \cmark&\cmark & \\
\hline
\multicolumn{1}{|c|}{Reference~\cite{COVID-DATA}} & 171 & 60 & 76 & \cmark &  &  &\cmark & \cmark & & \cmark&\cmark & \cmark\\
\hline
\end{tabular}
\footnotetext[1]{NA stands for Not Available.}
\end{adjustbox}
\vspace{-.1in}
\end{table*}
%%%%%%%%%%%%%%%%%%%%%%%%%%%%%%%%%%%
%%%%%%%%%%%%%%%%%%%%%%%%%%%%%%%%%%%
\begin{table*}[t!]
\centering
\caption{\label{tab:CXRdatasets}\textcolor{black}{Available COVID-19 CXR datasets.}}
\renewcommand{\arraystretch}{2}
\begin{adjustbox}{width=1\textwidth}
\begin{tabular}{c c|c|c|c|c|c|c|c|}
\cline{2-9}
 & \multicolumn{3}{|c|}{\textbf{Number of cases}}
 & \multicolumn{2}{|c|}{\textbf{Label type}}
 & \multicolumn{2}{|c|}{\textbf{Data Source}}
 & \multirow{2}{*}{{\textbf{Status}}}
 \\
\cline{1-8}
 \multicolumn{1}{|c|}{Dataset} & COVID & CAP & Normal  & Classification & Segmentation  & Multiple & Single& \\
\hline
\multicolumn{1}{|c|}{Reference~\cite{de2020bimcv}} & 802 & 605 & 284 & \cmark & \cmark  & \cmark & & Under development \\
\hline
\multicolumn{1}{|c|}{Reference~\cite{cohen2020covid}} & 468 & NA & NA & \cmark & \cmark  & \cmark & & Under development \\
\hline
\end{tabular}
\footnotetext[1]{NA stands for Not Available.}
\end{adjustbox}
\vspace{-.1in}
\end{table*}
%%%%%%%%%%%%%%%%%%%%%%%%%%%%%%%%%%%

\vspace{.05in}
\noindent
\textbf{\textit{COVID-19 CT Scans \& CXR Datasets:}} To assure model generalization for clinical use, it is beneficial to train SP/DL models based on a diverse set of dataset acquired from different scanners, different health centers covering a wide range of patients. Table~\ref{tab:datasets} provides an overview of the available CT imaging datasets along with their COVID-19 related information.
CT images represent different resolutions and contrasts depending on the type of scanner, image acquisition approach, and the thickness of the slices. It is, therefore, necessary to make CT images consistent before feeding them into the processing and learning models. For a list of available CT imaging datasets along with their COVID-19 related information please refer to Reference~\cite{COVID-DATA}.  Given the heterogeneity of the data source, available data collections comprehend a wide sort of equipment, images characteristics and diagnosed disease.

Table~\ref{tab:CXRdatasets} presents datasets of the two biggest data collection initiatives. Several other datasets are under development once when new data become available it is promptly aggregated by those initiatives. Given the heterogeneity of the data source, these data collection comprehends a wide sort of equipment, images characteristics and diagnosed disease.

%OOOOOOOOOOOOOOOOOOOOOOOOOOOOOOOOOOOOOOOOOOOOOOOOOOOOOOO
\section{Deep Learning Radiomics Specific to COVID-19} \label{sec:DR}
%OOOOOOOOOOOOOOOOOOOOOOOOOOOOOOOOOOOOOOOOOOOOOOOOOOOOOOO
In this section, we present different DL-based Radiomic models specific to the analysis of COVID-19 infection. In particular, we focus on the following  four different application domains of  discovery Radiomics: Segmentation of COVID-19 lesions (presented in Sub-section~\ref{subsec:Seg}); Predictive models for outcome prediction in COVID-19 patients (described in Sub-section~\ref{subsec:OP}); DL/SP models for severity assessment (presented in Sub-section~\ref{subsec:SA}), and; Diagnosis and classification of COVID-19 cases, which are detailed in Sub-section~\ref{subsec:Calss}.

%===================================================
\subsection{Segmentation of COVID-19 Lesions}\label{subsec:Seg}
%===================================================

\begin{figure*}[t!]
\centering
\includegraphics[width=0.65\textwidth]{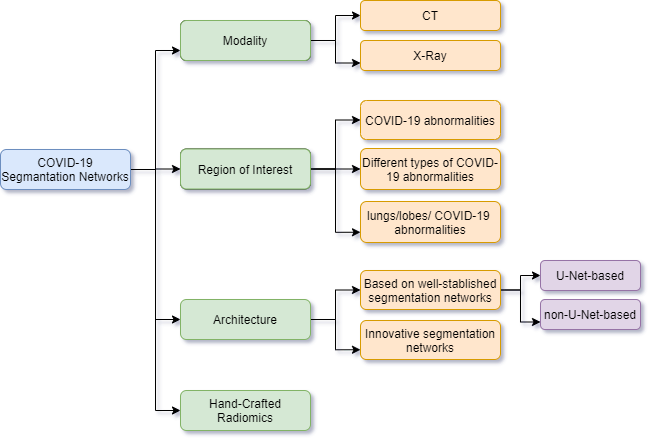}
\vspace{-.1in}
\caption{Taxonomy of the COVID-19 segmentation techniques using deep learning.}
\label{fig:tax_seg}
\vspace{-.2in}
\end{figure*}

In this sub-section, we provide an overview of segmentation networks developed in the context of COVID-19 from different aspects as presented in Fig.~\ref{fig:tax_seg}.
Segmentation networks are image-to-image DL models that are trained to produce a mask indicating the region of interest. Segmentation allows physicians to identify the type and location of lesions; Evaluate the extent of lung involvement, and; Quantify the lung severity measures. Depending on the objective, they one segment the COVID-19 lesions, lungs, and lobe regions. Segmentation-based infection quantification models can be used to evaluate effectiveness of different treatment solutions.

\vspace{.05in}
\noindent
\textit{\textbf{Imaging Modality used for Segmentation of COVID-19 Lesions:}} Since CT images provide the most accurate COVID-19 manifestations for grading and evaluation of infections, they have been widely used in the context of COVID-19 segmentation. The goal in this context is localization of COVID-19 infections and/or grading the disease stage. In the literature, the focus was mainly on development of 2D models for segmentation of lung infections in each CT slice~\cite{RajamaniSiebert, FanZhou, QiuLiu}. There have also been some 3D segmentation models that take the 3D CT volumes as input and segment the lung abnormalities on a patient-level basis~\cite{ZhangXiaohong, MullerRey}.
Since the use of portable CXRs is more feasible for patients in ICU, it is essential to develop segmentation models for severity assessment of COVID-19 patients based on CXR images.  A study on $2,951$ COVID-19 CXRs, performed the lung infections segmentation as the first step of their COVID-19 diagnosis pipeline~\cite{DegerliAhishali}. Using a human-machine collaboration, they provided the first COVID-19 CXR dataset with the ground-truth infection masks. Segmenting the lung infections using a U-Net model with DenseNet-121 yielded a higher performance in their classification framework~\cite{DegerliAhishali}.

\vspace{.05in}
\noindent
\textit{\textbf{Region of Interest (RoI):}} The region of interest would be different in COVID-19 segmentation models based on the research objective and can be classified into the following three main categories:
\begin{itemize}
\item [(i)] \textit{Localizing the Infection Regions without Considering their Types:} These studies perform a two-way segmentation approach, assigning each pixel of the CT images either to an infection or to a background class.
\item [(ii)] \textit{Segmenting Different Types of COVID-19 Lesions:} As mentioned previously, different types of COVID-19 infections can be correlated to the stage/severity of the disease. In this regard, some studies have segmented different types of COVID-19 lesions under different classes to further evaluate severity of the disease~\cite{ZhangXiaohong, RajamaniSiebert}.%, ZhengLiu}.
For instance, a 3D  DeepLabv3 segmentation model was proposed in~\cite{ZhangXiaohong} using CT images of $4,154$ patients. The constructed model segments lung regions and different types of COVID-19 infections, including consolidations, GGO, pulmonary fibrosis, interstitial thickening, and pleural effusion.
\item [(iii)] \textit{Segmentation of COVID-19 Lesions, Lungs, and Lobs:}  Researchers who aim to quantify the extent of lung involvement and determine the COVID-19 severity measures consider segmenting lung/lobe regions besides the COVID-19 lesions.
%%%%%%%%%%%%%%%%%%
%%%SEVERITY ASSESSMENT via CXR
%%%%%%%%%%%%%%%%%%
Advanced segmentation models can be trained to quantify different severity measures such as PO, PHO, CT score, and LHOS, based on CT images. This will be further discussed in Sub-section~\ref{subsec:SA}.
\end{itemize}
Next, we investigate different DL architectures proposed for the segmentation of COVID-19 abnormalities.
%%%%%%%%%%%%%%%%%%%%%%%%%%%%%%%%%%%%%%%
%%%%%%%%%%%%%%%%%%%%%%%%%%%%%%%%%%%%%%

%------------------------------------------------------------------------------------------------------
\subsubsection{DL Architectures for Segmentation of COVID-19 Lesions}
%------------------------------------------------------------------------------------------------------
Segmentation is considered an essential step in the severity/stage assessment of COVID-19 patients. However, comparing to classification models, limited number of research works have focused on segmentation models for COVID-19. Generally speaking, segmentation models (mostly developed based on CNNs) contain a contracting path (encoder) for extracting informative features from input images and an expanding path (decoder) for reconstructing the mask representing the regions of interest. Unlike classification models, typically, there are no fully connected layers in segmentation models. U-Net network~\cite{RonnebergerFischer}
and its various extensions are the most commonly used architecture for  segmentation of COVID-19 lesions. There are few works developed based on other successful segmentation networks. Some researchers have proposed innovative encoder-decoder networks for segmentation of COVID-19 lesions. Below, we present COVID-19 segmentation architectures classified into the aforementioned three main categories:
\begin{itemize}
\item [(ii)] \textbf{\textit{U-Net-based Segmentation Models:}} The majority of researches for  segmentation of COVID-19 abnormalities have been developed upon the U-Net model.  In U-Net~\cite{RonnebergerFischer},
skip connections transfer the extracted features from the contracting path to the corresponding layer in the expanding path. This helps the model to better understand visual details of images making it an ideal architecture for segmentation in the medical  domain. Adoption of pre-existing CNNs like DensNet and ResNet blocks in the encoder path of the U-Net will result in extracting higher resolution features from CT images~\cite{ChagantiGrenier, LiZhong}.  Multi-scale feature fusion, i.e., integration of dilated convolutions with different dilation rates, can be added into U-Net-based segmentation models to help capture COVID-19 abnormalities in different scales~\cite{WangLiu}. The current focus is on increasing performance in segmenting the COVID-19 lesions. In this regard, References~\cite{ZhouCanu, ZhengLiu, RajamaniSiebert}
considered incorporation of spatial and channel attention mechanisms within the U-Net architecture. The authors in Reference~\cite{RajamaniSiebert} proposed a new attention mechanism that enables the basic U-Net model to better understand the contextual information from CT slices and perform the segmentation of COVID-19 abnormalities more accurately. Commercial U-Net-based soft-wares have also been used in some researches for quantification of COVID-19 abnormalities and determining the severity of the disease~\cite{MergenKobe, HuangHan}.
\item [(ii)] \textbf{\textit{Non U-Net Segmentation Models:}} Some COVID-19 segmentation researchers developed models based on architectures other than the U-Net model. Fully Convolutional Networks (FCNs), U-Net++, and V-Net are some examples of successful segmentation networks that have been used as a base model for developing COVID-19 segmentation networks.
%Fully Convolutional Networks (FCNs)~\cite{LongShelhamer}, U-Net++~\cite{ZhouSiddiquee}, and V-Net~\cite{MilletariNavab} are some examples of successful segmentation networks that have been used as a base model for developing COVDI-19 segmentation networks. More specifically, Laradji, \textit{et al.}~\cite{LaradjiRodriguez} proposed a model for segmentation of COVID-19 lesions based on an FCN~\cite{LongShelhamer} network with VGG16 as the backbone. Shan, \textit{et al.}~\cite{ShanGao} used chest CT images of $249$ COVID-19 patients and developed a segmentation model based on VB-Net~\cite{MuLin}. VB-Net is a modification of V-Net~\cite{MilletariNavab} with bottleneck layers to decrease the number of feature map channels. Their segmentation model yielded a Dice score of $91.6$\% on a test set of $300$ COVID-19 patients. SCOAT-Net proposed in Reference~\cite{ZhaoLi} was developed based on U-Net++~\cite{ZhouSiddiquee} incorporating an attention learning mechanism. Their $5$-fold cross-validation results on a dataset containing $1,117$ annotated CT images from $19$ COVID-19 patients achieved a Dice score of $89.48$\% and a sensitivity of $88.74$\%.
%
\item [(iii)] \textbf{\textit{Innovative COVID-19 Specific Encoder-Decoders:}} Some studies have developed innovative models for the segmentation of COVID-19 opacifications from scratch. Qiu, \textit{et al.}~\cite{QiuLiu} proposed a compact (light-weight) segmentation network based on $100$ annotated CT slices from $>40$ COVID-19 patients. In contrary to most of the segmentation models, which are large DL networks with millions of trainable parameters, their proposed model contains only $472$ thousand parameters and achieved comparable results to its large-scale counterparts. Reference~\cite{FanZhou} developed a parallel partial decoder with edge and reverse attention modules to segment the COVID-19 lesions more precisely. Joint classification and segmentation networks, belonging to multi-task learning models, can enhance the performance of both segmentation and classification tasks by sharing the extracted features~\cite{WuGao, AmyarModzelewski}. Finally, segmentation models without use of labeled data can be developed~\cite{YaoXiao} for distinguishing COVID-19 areas of infections. More specifically, random shape, noise generation, and image filtering operations can be used to synthesize COVID-19 lesions for inclusion  into healthy chest CT scans and form training pairs.
\end{itemize}
Table~\ref{tab:SegModels} provides classifications of different COVID-19 lesion Segmentation models.

%------------------------------------------------------------------------------------------------------
\subsubsection{Hand-Crafted Radiomics}
\begin{figure}
\centering
\includegraphics[scale=0.2]{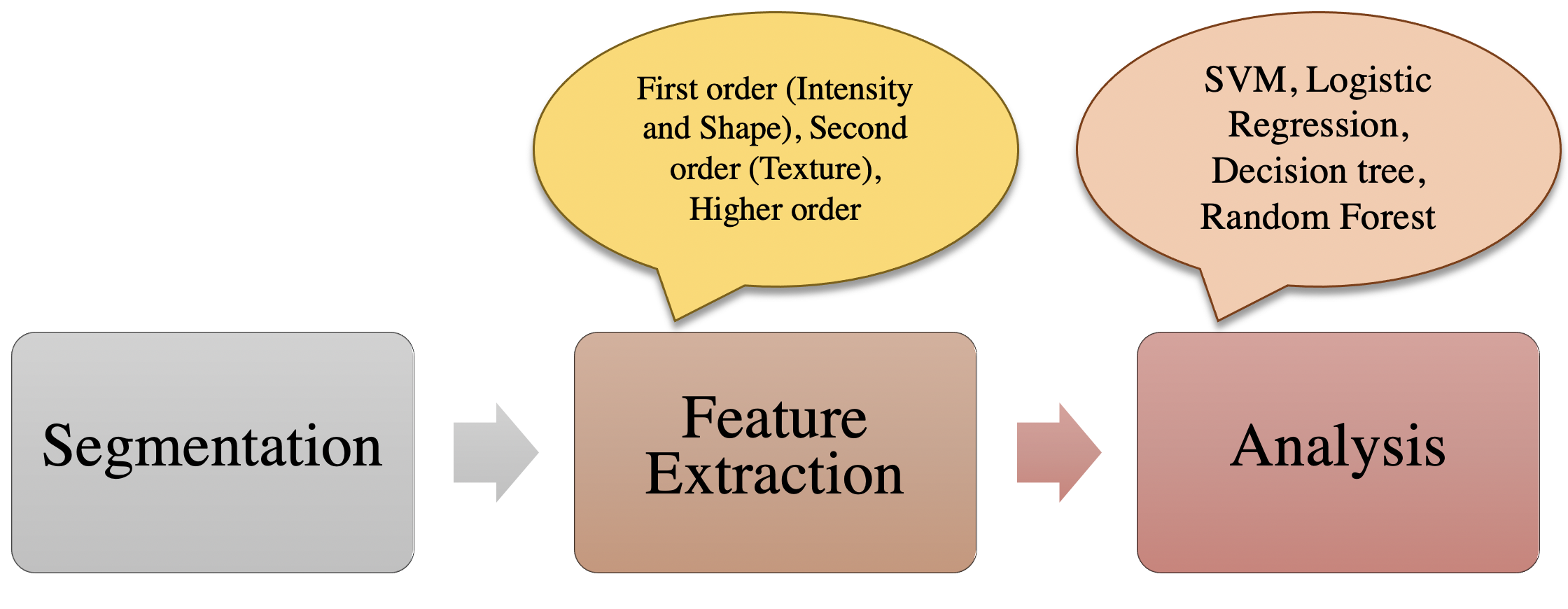}
\caption{Hand-crafted Radiomics workflow.}
\label{fig:hand-crafted}
\end{figure}

%------------------------------------------------------------------------------------------------------
Hand-crafted radiomics refers to the process of extracting several quantitative and semi-quantitative features from the ROI with the ultimate goal of diagnosis/prediction. Compared to DL techniques, Hand-crafted radiomics is less common in the problem of COVID-19 analysis, as it requires fine delineation of the infected regions and a prior knowledge of the types of the features to extract. Nevertheless, it benefits from more interpretability, as the features are engineered. As shown in Fig.~\ref{fig:hand-crafted}, Hand-crafted radiomics, utilized in a few COVID-19 studies, follows a multi-step process, in the first of which infected regions are annotated. Consequently, several features are extracted from the segmented regions and fed to a conventional model, such as Support Vector Machine (SVM), logistic regression, and decision tree, for making the final decision. Hand-crafted features cover a wide range of categories, including first-order (basic intensity and shape-based features), second-order (texture features extracted from various matrices), and more advanced features such as those calculated from Fourier and Wavelet transforms. Intensity features~\cite{ShiXia},
shape-based~\cite{FangHe},
and/or texture-based features, as well as other COVID-19 related features such as CT quantification metrics can be leveraged~\cite{LiZhong, chassagnon2020ai}. Radiomics in COVID-19 studies are mostly used in adverse outcome prediction models, explained in the next section. It is also possible to develop hybrid frameworks, where both hand-crafted and DL-based features are combined. Such methodology is adopted in Reference~\cite{WangWang} by combining GAN generated features with pre-defined hand-crafted ones.

%%%%%%%%%%%%%%%%%%%%%%%%%%%%%%%%%%%%%%
\onecolumn
 %\newgeometry{top=15mm, bottom=15mm, right=10mm, left=10mm}
 \begin{landscape}
 \begin{center}
 \centering
 %\fontsize{9}{11}\selectfont
 \begin{tiny}
 \begin{longtable}[c]{p{0.05\linewidth} p{0.05\linewidth} p{0.15\linewidth} p{0.1\linewidth} p{0.08\linewidth} p{0.1\linewidth} p{0.08\linewidth} p{0.08\linewidth} p{0.2\linewidth}}\\

 \caption{COVID-19 lesion Segmentation models}
 \label{tab:SegModels}\\

 \toprule
 \textbf{Ref.} & \textbf{Input data} & \textbf{Dataset size} & \textbf{Dataset diversity} & \textbf{Model} & \textbf{Task} & \textbf{ROI} & \textbf{Validation} & \textbf{Results}\\
 \midrule
 \endfirsthead

 \caption* {\textbf{Table \ref{tab:SegModels} Continued:} COVID-19 lesion Segmentation models}\\
 \toprule
 \textbf{Ref.}& \textbf{Input data} & \textbf{Dataset size} & \textbf{Dataset diversity} & \textbf{Model} & \textbf{Task} & \textbf{ROI} & \textbf{Validation} & \textbf{Results}\\
 \midrule
 \endhead
 Ref.~\cite{ChagantiGrenier}& CT scans& 9749 CT volumes& multi-center& U-Net-based& Quantification&lesions, Lungs, lobes& Train/test split& Pearson correlation for PO: 0.92, PHO: 0.97, LSS: 0.91, LHOS: 0.9\\
 \\
 Ref.~\cite{RajamaniSiebert}& CT scans & 471 slices (100 slice from 40 patients, 371 slices from 9 patients) & multi-center & U-Net-based & Segmentation & COVID-19 lesions (binary/multiple class segmentation) & 3-fold cross validation & DSC: 0.791, SEN: 0.862, SPC: 0.987\\
 \\
 Ref.~\cite{MaWang}& CT scans&	20 COVID-19 CT volumes from 2 datsets (10 patients from each)& multi-center&	U-Net-based&	Segmentation&	COVID-19 lesions, Lungs&	4-fold cross validation&	DSC: on left lung: 0.8581, DSC on right lung: 0.8799, DSC on lesions: 0.6732\\
 \\
 Ref.~\cite{LaradjiRodriguez}& CT scans&	29 COVID-19 CT volumes& multi-center&	FCN-based&	Segmentation&	COVID-19 lesions&	Train/test split&	Reducing annotating time \\
 \\
 Ref.~\cite{YaoXiao}& CT scans&	For COVID-19 lesion segmentation: (453 healthy CT volumes, 18 COVID-19 CT volumes) For lung segmentation: 515 lung CT volumes&	Three datasets for COVID-19 lesion segmentation, Three datasets for lung segmentation&	U-Net-based&	Segmentation&	COVID-19 lesions, Lungs&	Eternal validation&	Results on test set 1: DSC: 0.687 $\pm$ 15.8, SPC: 0.851 $\pm$ 6.97, SEN: 0.621 $\pm$ 22.8 ; Results of test set 2:  DSC: 0.594 $\pm$ 17.4, SPC: 0.604 $\pm$ 19.7, SEN: 0.618 $\pm$ 18.4 \\
 \\
 Ref.~\cite{XuCao}& CT scans&	2563 COVID-19 CT volumes, 254 healthy CT volumes&	four public datasets for training and three public datasets for testing&	U-Net-based + GAN&	Segmentation&	COVID-19 lesions&	External validation&	Results on different test sets: DSC: 0.589-0.767, SPC: 0.992-0.998, SEN: 0.584-0.846\\
 \\
 Ref.~\cite{QiuLiu}& CT scans&	100 annotated CT slices form $>40$ patients&	single-center&	Innovative model (light-weight architecture)&	Segmentation&	COVID-19 lesions&	Train/test split&	DSC: 0.7728, SEN: 0.8362, SPC: 0.9747\\
 \\
 Ref.~\cite{AmyarModzelewski}& CT scans&	1396 CT slices for classification, 100 slices from $>40$ patients for segmentation&	Three datasets for classification, one dataset for segmentation&	Innovative model (multi-task learning)&	Segmentation + Classification +Image reconstruction&	COVID-19 lesions&	Train/test split&	For segmentation DSC: 0.88; For classification Acc: 0.9467, SEN: 0.96, SPC: 0.92\\
 \\
 Ref.~\cite{HuangHan}& CT scans&	The commercial deep-learning software used for segmentation has been trained based on 842 COVID-19 CT volumes from one hospital. For follow up CTs: 126 COVID-19 patients classified into four clinical stages: 6 mild, 94 moderate, 20 severe, and 6 critical cases&	Single-center&	U-Net-based&	Quantification of lung involvement, monitoring changes in follow up CTs&	COVID-19 lesions, Lungs, Lobes&	Train/test split&	DSC: median: 0.8481, range: 0.6526 - 0.9094; PO for different categories: mild: 0, moderate: 2.2\% (0.4, 7.1), severer: 28.9\%  $\pm$  19.2, critical: 49.6\%  $\pm$  14.8\\
 \\
 Ref.~\cite{FanZhou}& CT scans&	100 annotated CT slices from $>40$ patients, 1600 CT slices from 20 patients&	multi-center&	Innovative model&	Segmentation&	COVID-19 lesions (Bianry/multiple class segmentation)&	External validation& Results on test set: DSC: 0.73, SEN: 0.725, SPC:  0.96, Results on external validation: DSC: 0.597, SEN: 0.865, SPC: 0.977\\
 \\
 Ref.~\cite{ZhouCanu}& CT scans&	100 annotated CT slices from $>40$ patients& 	multi-center&	U-Net-based&	Segmentation&	COVID-19 lesions&	Train/test split&	DSC: 0.83.1 , SEN: 0.867, SPC: 0.993\\
 \\
 Ref.~\cite{WangLiu}& CT scans&	558 COVID-19 patients including 76250 slices& 	multi-center&	U-Net-based&	Segmentation&	COVID-19 lesions&	Train/test split&	DSC: 0.8029  $\pm$  11.14, HD:18.72  $\pm$  27.26\\
 \\
 Ref.~\cite{ZhaoLi}& CT scans&	1117 annotated CT images from 19 COVID-19 patients, for external validation: 8 lung CT scans from two COVID-19 patients&	single-center&	U-Net++-based&	Segmentation&	COVID-19 lesions&	5-fold cross-validation&	Results on test set: DSC: 0.8948, SEN: 0.8874, PPV: 0.9064; Results on external set: only qualitative results have been presented\\
 \\
 Ref.~\cite{LiZhong}& CT scans&	531 thick-section CT scans from 204 COVID-19 patients&	single-center&	U-Net-based&	Segmentation; Severity assessment using PO and the average infection HU (iHU)&	COVID-19 lesions&	Train/test split&	For segmentation: DSC: 0.74, For severity assessment: two imaging bio-markers (PO and iHU) can distinguish between the severe and non-severe stages with an AUC of 0.9680 (p-value$< 0:001$).\\
 \\
 Ref.~\cite{MergenKobe}& CT scans, clinical/laboratory information& 	60 COVID-19 patients&	single-center&	U-Net-based&	abnormalities quantification, correlation with disease severity&	COVID-19 lesions/ Lungs/ Lobes&	NA&	Patients with need for mechanical ventilation had a significantly higher PO (median 44\%, IQR: 23–58\% versus 13\%, IQR: 10–24\%; p = 0.001) and PHO (median: 11 \%, IQR: 6–21\% versus 3\%, IQR: 2–7 \%, p = 0.002) compared to those without.\\
 \\
 Ref.~\cite{MaNie}& CT scans&	880 COVID-19 CT volumes, 80 annotated cases&	multi-center&	nnU-Net-based&	Segmentation&	COVID-19 lesions&	5-fold cross validation&	DSC: 0.7225 $\pm$ 0.1989, HD: 23.46 $\pm$  32.12\\
 \\
 Ref.~\cite{ZhengLiu}& CT scans&	2506 slices with COVID-19 infected lesions and 2274 non-infected slices from 18 COVID-19 patients and 18 non-COVID people&	single-center&	U-Net-based&	Segmentation&	COVID-19 lesions (including GGOs, interstitial in ltrates, and consolidation)&	5-fold cross validation&	for three infection categories: DSC:  0.7422,0.7384,0.8769.SEN and SPC: (0.8593, 0.9742), (0.8268,0.9869) and (0.8645,0.9889) respectively\\
 \\
 Ref.~\cite{ZhangXiaohong}& CT scans,CT quantitative feature,clinical data&	617,775 CT images from 4,154 patients, 4695 annotated slices&	multi-center& 	DeepLabv3-based& Three-way classification, Segmentation, quantification& Lung regions, COVID-19 lesions including consolidation, GGO, pulmonary fibrosis, interstitial thickening, and pleural effusion&	5-fold cross validation, External validation& three-way classification: accuracy: 0.9249, AUROC: 0.9813 (95\% CI: 0.9691–0.9902); The classification performance on five external datasets:  accuracy: above 0.8411, sensitivity above 0.8667, and  specificity above 0.8226; The AI-based lesion quantification could evaluate the effectiveness of different treatment solutions"\\
 \\
 Ref.~\cite{WuGao}& CT scans& 750 CT volumes (400 COVID-19 \& 350 uninfected cases). 3,855 annotated CT slices&	multi-center&	Innovative model&	binary-classification, Segmentation&	COVID-19 lesions&	Train/test split&	Classification: sensitivity: 0.95, specificity: 0.93; segmentation: DSC: 0.783\\
 \\
 Ref.~\cite{ShanGao}& CT scans&	549 COVID-19 patients &	single-center&	VB-Net-based&	Segmentation, quantification& Covid-19 lesions, Lungs, Lobes&	Train/test split& DSC: 0.916\%  $\pm$  10, PO error (the difference between GT and predicted PO): 0.3\%\\
 \hline

 \end{longtable}
 \end{tiny}
 \end{center}
 \end{landscape}
% \restoregeometry
\twocolumn
%%%%%%%%%%%%%%%%%%%%%%%%%%%%%%%%%%%%%

%%%%%%%%%%%%%%%%%%%%%%%%%%%%%%%%%%%%%%%
%%%%%%%%%%%%%%%%%%%%%%%%%%%%%%%%%%%%%%

%%%%%%%%%%%%%%%%%%%%%%%%%%%%%%%%%%
%\begin{landscape}
 \begin{table*}[t!]
 \centering
 \caption{\label{tab:predictModels}\textcolor{black}{Predictive models for outcome prediction in COVID-19 patients.}}
 \renewcommand{\arraystretch}{2}
 \begin{adjustbox}{width=\textwidth}
 \begin{tabular}{c p{0.1\linewidth}|p{0.1\linewidth}|p{0.25\linewidth}|p{0.25\linewidth}|p{0.15\linewidth}|p{0.15\linewidth}|p{0.15\linewidth}|p{0.15\linewidth}|p{0.2\linewidth}|p{0.2\linewidth}|p{0.2\linewidth}|p{0.2\linewidth}|}

  \\
 \hline
  \multicolumn{1}{|c|}{\textbf{Reference}} & \textbf{Dataset Size} & \textbf{Dataset Diversity}  & \textbf{Input Data} & \textbf{Model}  & \textbf{Target Outcome} \\
 \hline
 \multicolumn{1}{|c|}{Ref.~\cite{yue2020machine}} & 31 patients & Multi-center & extracted CT features & Logistic regression/ Random forest & short- and long-term hospital stay \\
 \hline
 \multicolumn{1}{|c|}{Ref.~\cite{bai2020predicting}} & 133 patients in mild stage &  single-center & Temporal information of CT scans and clinical/laboratory data & A joint multi-layer perceptron and LSTM network & Progression from mild stage to sever/critical stage \\
 \hline
 \multicolumn{1}{|c|}{Ref.~\cite{zeng2020risk}} & 338 patients & Single-center & Extracted features from CT images/ clinical variables & Multivariate survival analyses & Progression Risk \\
 \hline
 \multicolumn{1}{|c|}{Ref.~\cite{colombi2020well}} & 236 patients & Single-center & clinical parameters and CT metrics & Logistic regression &  ICU admission or death vs no ICU admission or death \\
 \hline
 \multicolumn{1}{|c|}{Ref.~\cite{chassagnon2020ai}} & 693 patients & multi-center & Radiomic CT features, clinical/ biological attributes & Ensemble consensus-driven learning  &  Severe vs non-severe/ short vs long-term prognosis\\
 \hline
 \multicolumn{1}{|c|}{Ref.~\cite{lassau2020ai}} & 1003 patients & multi-center & Clinical, biological, and CT scan images and reports & DL pipeline for segmentation/ DL pipeline for predicting severity evolution & Progression Risk \\
 \hline
 \end{tabular}
 \end{adjustbox}
 \end{table*}
%\end{landscape}
%%%%%%%%%%%%%%%%%%%%%%%%%%%%%%%%%%%%%

%===================================================
\subsection{Predictive Models for Adverse Outcome Prediction in COVID-19 Patients} \label{subsec:OP}
%===================================================
As stated previously, for efficient utilization of limited medical resources during the COVID-19 pandemic, it is critically important to accurately predict mortality risk, progression to severe/critical stage, need for ICU admission/ventilation, and the length of hospital stay. In this regard, development of predictive models is essential to compute the probability of poor outcomes and help health-care professionals deliver appropriate services to high-risk patients.

\begin{figure*}
\centering
\includegraphics[scale=0.6]{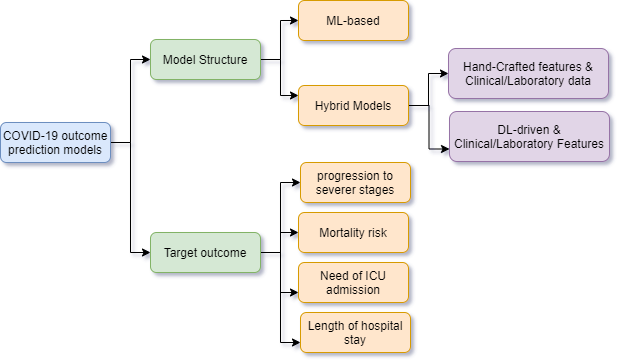}
\caption{Taxonomy of the COVID-19 predictive models.}
\label{fig:outcome}
\end{figure*}

 Although image-driven features have shown high correlation with COVID-19 outcomes, they are not the only influential factors. In other words, radiologists use image-driven features together with other clinical and risk factors to make the final decision. Some of the clinical/laboratory information used for COVID-19 outcome prediction are patients' symptoms, laboratory test results, oxygen saturation, and comorbid diseases. Chronic lung disease, obesity, hypertension, cardiovascular diseases, and diabetes are examples of comorbidities that will increase the risk of adverse outcomes in COVID-19 pneumonia. In this section, we focus on predictive models that image-driven features to estimate the risk of adverse outcomes in COVID-19 patients. As shown in Fig.~\ref{fig:outcome}, in what follows, predictive models are categorized and described in terms of their: (i) Model structure, and; (ii) Target outcome.

%-------------------------------------------------------------------------------------------------------
\subsubsection{Model Structure of COVID-19 Outcome Prediction Models}
%-------------------------------------------------------------------------------------------------------
%%%%%%%%%%%%%%%%%%%%%%%%%%%%%%%%%%%%%%%
%%%%%%%%%%%%%%%%%%%%%%%%%%%%%%%%%%%%%%%
Most COVID-19 outcome prediction studies exploit both chest CT images and clinical/laboratory data in their models. To effectively benefit from heterogenous data resources,   conventional ML methods or hybrid models can be utilized, as explained below.

\vspace{.1in}
\noindent
(i) \textbf{\textit{Conventional ML Models:}} In some COVID-19 predictive studies, CT Radiomics and quantification features are extracted in a pre-processing step. Extracted features are then used with clinical/laboratory data to train a shallow classifier such as logistic regression or random forest. For instant, Chao \textit{et al.}~\cite{chao2020integrative} used CT features including lobe-wise quantification features and whole lung Radiomics together with patients' clinical information including age, sex, vital signs, and laboratory findings to predict the need for ICU admission in COVID-19 patients. They used a DL-based segmentation model to measure CT quantification features. Integrated input data from various types and resources are then fed into a random forest classifier for outcome prediction. Following this study, one can conclude that adding clinical information to CT features can improve the overall outcome prediction performance.

\vspace{.1in}
\noindent
(ii) \textbf{\textit{Hybrid Models:}}  Hybrid models (such as multiple-models, mixture of experts, and ensemble models) are of high importance  in the field of medical imaging, typically, improving the initial results. While hybrid models can be developed in a variety of forms, they are mostly adopted in COVID-19 analysis in two main ways, i.e., combinations of Hand-Crafted and Clinical/Laboratory features or combination of DL-driven and clinical/Laboratory features, as described below:
\begin{itemize}
\item \textit{Combinations of Hand-Crafted Features and Clinical/Laboratory Information:}  Although hand-crafted Radiomics, typically, rely on fine annotations and prior knowledge of the features to be extracted, they benefit from domain knowledge. Combining the CT Radiomics and clinical/laboratory information in a hybrid model, therefore, benefits from more interpretability. After integration of hand-crafted Radiomics and clinical/laboratory information, DL features can be extracted for achieving improved overall results.  For instance, DL models can be developed to predict the probability that mild COVID-19 patients deteriorate into the severe/critical stage. In this regard, Reference~\cite{bai2020predicting} developed a DL model to predict the probability that mild COVID-19 patients deteriorate into the severe/critical stage. In their model, first, clinical/laboratory data are fed into a Multi-Layer Perceptron (MLP). The output is then integrated with extracted hand-crafted features from serial CT scans, which are then fed into an LSTM network followed by fully connected layers. The LSTM network could detect the temporal dependencies between the vector of features and achieved an AUC of $0.92$ in distinguishing mild patients who are more likely to deteriorate into the severe/critical stage.
\item \textit{Combination of DL-driven and Clinical/Laboratory Features}:
 The superiority of a mixture model that takes advantage of both image-driven and clinical factors is investigated in~\cite{MengDong,NingLei}.
In this study, gender, age, severity grade, and chronic disease history are combined with the chest CT scans through a joint CNN-MLP network to distinguish between high and low-risk COVID-19 patients. In another study~\cite{NingLei},
 blood and urine test results of $1,170$ patients are used as the clinical information. The developed model consists of two successive CNN networks for analysis of CT images, i.e., a DNN network for analysis of clinical features, and a penalized logistic regression to integrate image-driven DL features and the DL features extracted from clinical data. Improvements are reported when image-driven and clinical features are jointly used.
\end{itemize}

%----------------------------------------------------------------------------------------------
\subsubsection{Target Outcomes in COVID-19 Patients}
%----------------------------------------------------------------------------------------------
%%%%%%%%%%%%%%%%%%%%%%%%%%%%%%%%%%%%%%%
\begin{figure*}[t!]
\centering
\centering
\includegraphics[scale=0.5]{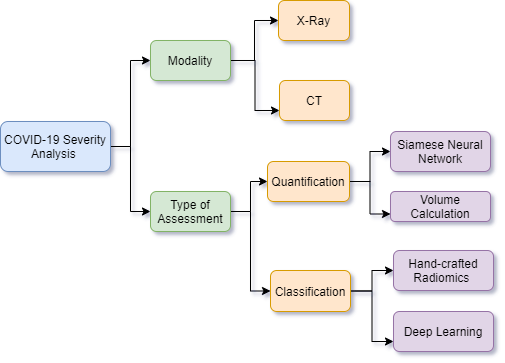}
\vspace{-.15in}
\caption{Taxonomy of the COVID-19 severity assessment techniques using deep learning.}\label{fig:tax_severity}
\vspace{-.2in}
\end{figure*}
%%%%%%%%%%%%%%%%%%%%%%%%%%%%%%%%%%%%%%%
Generally speaking, in the context of COVID-19 prognosis, the target outcomes of interest include progression to severe/critical stage, mortality risk, need for ICU admission/ventilation, and the length of hospital stay. Below, we present different COVID-19 outcome prediction models based on these target outcomes:
\begin{itemize}
\item \textit{Risk of Progression to Severer Stages:} Li \textit{et al.}~\cite{LiZhong} aimed at measuring the progression of the disease by monitoring the Portion of Infection (POI) and the average infection HU (iHU). Changes in POI and iHU are calculated over each two consecutive CT examinations and compared with the radiologists' reports, leading to a high agreement in determining the infection increase or decrease. To understand the temporal evolution of COVID-19, Reference~\cite{HuangHan} calculates the Lung Opacity Percentage (LOP) for the whole lung and its $5$ lobes over follow-up CT examinations. Progression assessment has also been followed in Reference~\cite{AmerFrid-Adar} by calculating the pneumonia ratio.
\item \textit{Mortality Risk:} Chassagnon \textit{et al.}~\cite{chassagnon2020ai} proposed an ensemble consensus-driven learning approach with a combination of multiple ML classifiers. They used CT Radiomic features (including first-order, higher-order statistics, texture, and shape information) and patients' data (including age and gender) as the  model's input. Their model aimed to predict a short term negative outcome (death in less than four days) or a long-term negative outcome (patients who did not recover after $31$ days may die after four days or still intubated). They trained their model based on a multi-center dataset containing $693$ COVID-19 patients and obtained promising results on multiple external validation sets.
\item \textit{Need of ICU Admission:} Reference~\cite{colombi2020well} implemented a
logistic regression model to predict the risk of ICU admission/death based on clinical parameters and CT quantification metrics of 236 COVID-19 patients from one health center.
CT quantification metrics, i.e., the extent of lung involvement by COVID-19 infections, can also be used to predict ICU admission or death. PO and PHO are calculated in Reference~\cite{MergenKobe} for potential correlation with clinical and laboratory factors. Obtained results show that patients with high PO and PHO have higher needs for mechanical ventilation. Length of ICU stay, the duration of oxygen inhalation, and hospitalization are the main follow up objectives in Reference~\cite{CaiLiu}. Homayounieh \textit{et al.} \cite{HomayouniehEbrahimian} used a logistic regression model to predict the risk of ICU admission or death based on CT radiomics and clinical information. They indicated that CT radiomics could be superior to radiologists' visual assessment in predicting COVID-19 patients' outcomes.
\item \textit{Length of Hospital Stay:} Need for short or long-term hospital stay of COVID-19 patients can be estimated using regression or random forest models. In such scenarios, hand-crafted features such as first-order, second-order, and/or shape features are extracted to predict the length of hospital stay of COVID-19 patients. Yue \textit{et al.}~\cite{yue2020machine} estimated the length of hospital stay in COVID-19 patients using logistic regression and random forest techniques.
\end{itemize}

%===================================================
\subsection{DL/SP Models for COVID-19 Severity Assessment}\label{subsec:SA}
%===================================================
Severity essentially refers to how much the lungs are affected and involved in the disease. COVID-19 severity assessment is of high importance due to its unique role in risk management and resource allocation. In this sub-section, as shown in Fig.~\ref{fig:tax_severity}, we present existing severity assessment methodologies. Table~\ref{tab:SevirityModels} summarizes how different studies are categorized based on these perspectives.

\vspace{.05in}
\noindent
\textit{\textbf{Imaging Modality used for Severity Assessment of COVID-19:}}  Severity of the COVID-19 can be assessed using both CXR and CT scans, where the latter, due to its 3D nature, is capable of providing more accurate estimate of the lung involvement. Below, we provide few examples of recent works using CXR or CT for severity assessment of COVID-19 infection:
\begin{itemize}
\item \textit{CXR for Severity Assessment}: %Reference~\cite{ZhuShen}
For utilization of CXR for severity assessment, irrelevant, low quality and negative COVID-19 images need to be excluded prior to the analysis~\cite{ZhuShen}. CXR are also used in Reference~\cite{AmerFrid-Adar} for severity assessment.
\item \textit{CT for Severity Assessment}: RT-PCR-confirmed CT scans are utilized in Reference~\cite{MergenKobe}  for severity assessment, accompanied by clinical and laboratory data, as well as the need for oxygen supply and mechanical ventilation. CT scans are also utilized by Li \textit{et al.}~\cite{LiZhong} and divided into severe and non-severe groups. The non-severe cases may progress into the severe class during the treatment. Since severe and non-severe patients have different treatment regimens, the same grouping is performed in Reference~\cite{TangZhao}. CT scans from multiple centers are utilized by Ghosh \textit{et al.}~\cite{GhoshKumar} for a generalizable severity assessment. As stated previously, segmentation models are, typically, used to quantify different severity measures such as PO, PHO, CT score, and LHOS, based on CT images. More specifically, to quantify (PO, CT score) and (PHO, LHOS), the model learns to segment COVID-19 infections and COVID-19 high-opacity infections, respectively. The whole lung region and lobe regions are segmented to measure (PO, PHO) and (CT score, LHOS).  For instance, Reference~\cite{HuangHan}, segmented the lungs regions and COVID-19 lesions using a U-Net-based commercial software and determined the PO in COVID-19 patients. Based on the clinical data, they could map the PO measure to the severity of the disease. It was concluded in this study that the median PO for patients in the moderate category is $2.2$ ($0.4$, $7.1$), for patients in the severe group is $28.9$ $\pm$ $19.2$, and for patients in the critical category is $49.6$ $\pm$ $14.8$. Along the similar path, Reference~\cite{ShanGao} developed a VB-Net-based segmentation model using $249$ CT volumes of COVID-19 patients to segment the lung regions, lobes, and lung infections. Their model could quantify the PO measure with an error of $0.3$\% over a testset of $300$ CT volumes collected at a single hospital. Commercial U-Net-based soft-wares have also been used in some researches for quantification of COVID-19 abnormalities and determining the severity of the disease~\cite{MergenKobe, HuangHan}.
\end{itemize}
%
%%%%%%%%%%%%%%%%%%%%%%%%%%%%%%%%%%%%%%
\onecolumn
% \newgeometry{top=15mm, bottom=15mm, right=10mm, left=10mm}
 \begin{landscape}
 \begin{center}
 \centering
 %\fontsize{9}{11}\selectfont
 \begin{footnotesize}
 \begin{longtable}[c]{p{0.05\linewidth} p{0.05\linewidth} p{0.15\linewidth} p{0.1\linewidth} p{0.08\linewidth} p{0.1\linewidth} p{0.15\linewidth} p{0.1\linewidth}}\\

 \caption{COVID-19 severity assessment models.}
 \label{tab:SevirityModels}\\

 \toprule
 \textbf{Ref.} & \textbf{Input data} & \textbf{Dataset size} & \textbf{Dataset diversity} & \textbf{Objective} & \textbf{Segmentation method} & \textbf{Type of assessment}\\
 \midrule
 \endfirsthead

 \caption* {\textbf{Table \ref{tab:ClassModels} Continued:} COVID-19 severity assessment models.}\\
 \toprule
 \textbf{Ref.} & \textbf{Input data} & \textbf{Dataset size} & \textbf{Dataset diversity} & \textbf{Objective} & \textbf{Segmentation method} & \textbf{Type of assessment}\\
 \midrule
 \endhead
 Ref.~\cite{GhoshKumar} & CT & 509 CT images from 101 COVID-19 patients & Multi-center & Diagnosis & Deep learning; Traditional; Manual & Hand-crafted radiomics\\
 \\% LyuLiu ZhuShen AmerFrid-Adar YipKlanecek FengLiu CaiLiu LiArun
 Ref.~\cite{LyuLiu} & CT & 51 patients & One hospital & Diagnosis & Automatic followed by manual adjustment & Volume calculation\\
 \\
 Ref.~\cite{HuangHan} & CT & 842 COVID-19 CT volumes for segmentation; 126 COVID-19 patients classified into four clinical stages: 6 mild, 94 moderate, 20 severe, and 6 critical cases & One hospital & Progression assessment & Deep learning & Volume calculation\\
 \\
 Ref.~\cite{LiZhong} & CT & 531 thick-section CT scans from 204 COVID-19 patients & One hospital & Progression assessment & Deep learning & Volume calculation\\
 \\
 Ref.~\cite{MergenKobe} & CT & 60 COVID-19 patients & One hospital & Correlation with clinical factors & Deep learning & Volume calculation\\
 \\
 Ref.~\cite{ZhuShen} & X-ray & 131 portable CXR from 84 COVID-19 patients & One public dataset & Diagnosis & Not required & Deep learning\\
 \\
 Ref.~\cite{TangZhao} & CT & 176 patients & 7 hospitals with different scanners & Diagnosis & Deep learning & Volume calculation\\
 \\
 Ref.~\cite{AmerFrid-Adar} & X-ray & 15756 COVID-19 images & Several public datasets & Progression assessment & Deep learning & Volume calculation\\
 \\
 Ref.~\cite{YipKlanecek} & CT & 1110 COVID-19 patients  & One public dataset & Diagnosis & Traditional & Hand-crafted radiomics\\
 \\
 Ref.~\cite{FengLiu} & CT & 346 COVID-19 patients & Two hospitals &Progression assessment & Deep learning & Deep learning\\
 \\
 Ref.~\cite{CaiLiu} & CT & 99 COVID-19 patients & Two institutions & Correlation with clinical factors & Deep learning & Hand-crafted radiomics\\
 \\
 Ref.~\cite{LiArun} & X-ray & 468 COVID-19 patients & One hospital & Progression assessment & Not required & Siamese neural networks\\
 \\
 \hline

 \end{longtable}
 \end{footnotesize}
 \end{center}
 \end{landscape}
% \restoregeometry
\twocolumn
%%%%%%%%%%%%%%%%%%%%%%%%%%%%%%%%%%%%%%
%------------------------------------------------------------------------------------------------------
\subsubsection{Severity Assessment Types}
%------------------------------------------------------------------------------------------------------
%%%%%%%%%%%%%%%%%%%%%%%%%%%%%%%%%%%%%%%
\begin{figure}[t!]
\centering
\includegraphics[scale=0.4]{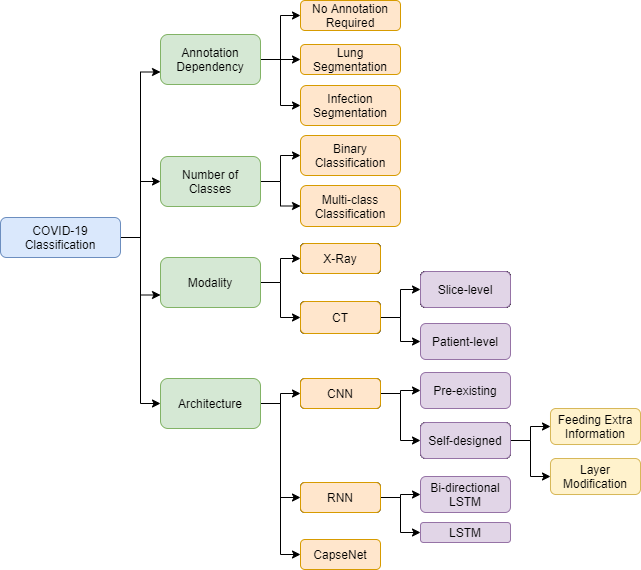}
\caption{Taxonomy of DL-based COVID-19 classification techniques.}\label{fig:tax_class}
\vspace{-.2in}
\end{figure}
%%%%%%%%%%%%%%%%%%%%%%%%%%%%%%%%%%%%%%%
Generally speaking, two types of severity assessment can be defined within the COVID-19 literature. The first one is to consider a classification approach, where different discrete labels are defined to assess the severity. The second type, however, aims at calculating the degree/portion of lung involvement as a measure of severity. Although, the second type, referred to as ``\textit{quantification}'', is often followed by a classification paradigm, degree of lung involvement is essentially embedded in the feature vector. Below, we further elaborate on these two severity assessment types:

\vspace{.1in}
\noindent
(i) \textit{\textbf{COVID-19 Severity Classification:}} Similar to most of the classification problems, COVID-19 severity classification can be solved using either hand-crafted or DL methods.
\begin{itemize}
\item \textit{Hand-crafted Radiomics}: While different engineered features have the potential to distinguish between severe and non-severe cases, Ghosh \textit{et al.}~\cite{GhoshKumar} proposed a hand-crafted feature, referred to as the $L_{norm}$. This feature is defined using the maximum bone reference ($B$), minimum air reference ($A$), and the mean gray scale intensity of the lesion ($L$), as follows
\begin{equation}
L_{norm}=100\times\frac{L-A}{B-A},\quad 0\leq L_{norm} \leq 100.
\end{equation}
The optimum cut-off value to distinguish between severe and non-severe cases using the $L_{norm}$ is then obtained based on a Receiver Operating Characteristic (ROC) curve analysis. Other traditional hand-crafted features, such as first-order histogram features and/or texture-based ones can be incorporated followed by a regression model to distinguish between severe and non-severe patients. First-order histogram features are also used in Reference~\cite{CaiLiu} for severity classification.
\item \textit{Deep Learning}: To identify discrete severity scores of the COVID-19 patients, CNN-based models can alternatively developed~\cite{ZhuShen}. A two stage DL framework is proposed in Reference~\cite{FengLiu} for COVID-19 severity classification. In the first stage, CT scans are individually fed to a U-Net model, whose extracted features are stored for the second stage. Through the second stage, the feature vectors are fed to a bi-directional LSTM model, for the final classification.
\end{itemize}

\vspace{.1in}
\noindent
(ii) \textit{\textbf{Severity Assessment via Quantification:}} Although quantification is performed by calculating the lung and infection volume in most of the studies, it is also possible to adopt a different approach such as using a Siamese neural network. Below, we discuss recent works performed along these two directions:
\begin{itemize}
\item \textit{Quantification via Volume Calculation}: To quantify the COVID-19 severity, Reference~\cite{MergenKobe} calculated PO and PHO.  POI and the average infection HU are calculated in Reference~\cite{LiZhong}, to quantify the severity, followed by dividing the patients into two groups of severe and non-severe. Infection and GGO ratio are calculated in Reference~\cite{TangZhao}. These two measures, along with several other quantitative features, are further fed to a Random Forest (RF) classifier, to classify patients as severe and non-severe.
\item \textit{Quantification via Siamese Neural Networks}: This approach consists of two identical models, in terms of weights and parameters, with the goal  of finding the similarity between the two inputs. Beside having several applications, Siamese models, in particular convolutional Siamese neural networks, can be adopted for COVID-19 severity assessment~\cite{LiArun}
In such scenarios, the Euclidean distance between the two final layers is calculated as a measure of difference between the inputs. Therefore, the distance between a COVID-19 and normal scan can show the degree of abnormality. Utilizing a pool of normal images, the median distance can represent the severity.
\end{itemize}

%===================================================
\subsection{COVID-19 Classification Models}\label{subsec:Calss}
%===================================================

Development of DL-based COVID-19 classification models can be approached from four main perspectives, as shown in Fig.~\ref{fig:tax_class}. The first aspect is the annotation dependency of the developed frameworks. The second one is whether the proposed methods consider a binary or multi-class classification, followed by the third perspective focusing on the imaging modality used for the classification, as based on the modality different solutions are admissible. The DL model architecture is another important aspect of the developed frameworks. Next, we discuss these four categories in detail.

Table~\ref{tab:ClassModels} summarizes how different studies approach the aforementioned categories.

%------------------------------------------------------------------------------------------------------
\subsubsection{Annotation Dependency}
%------------------------------------------------------------------------------------------------------
Annotation dependency refers to whether the developed COVID-19 classification models rely on annotated images as inputs. Annotation can be related to either segmenting the whole lung region or the infected areas from the chest image. In this regard, we categorized studies into three groups of (i) no annotation required, (ii) lung segmentation required, and (iii) infection segmentation required. These three groups are discussed in the following:

%%%%%%%%%%%%%%%%%%%%%%%%%%%%%%%%%%%%%%
\onecolumn
% \newgeometry{top=15mm, bottom=15mm, right=10mm, left=10mm}
 \begin{landscape}[t!]
 \begin{center}
 \centering
 %\fontsize{9}{11}\selectfont
 \begin{tiny}
 \begin{longtable}[c]{p{0.05\linewidth} p{0.05\linewidth} p{0.15\linewidth} p{0.1\linewidth} p{0.08\linewidth} p{0.1\linewidth} p{0.15\linewidth} p{0.1\linewidth}}\\

 \caption{COVID-19 classification models.}
 \label{tab:ClassModels}\\

 \toprule
 \textbf{Ref.} & \textbf{Input data} & \textbf{Dataset size} & \textbf{Dataset diversity} & \textbf{Number of classes} & \textbf{Architecture} & \textbf{Bias and over-fitting prevention} & \textbf{Annotation dependency}\\
 \midrule
 \endfirsthead

 \caption* {\textbf{Table \ref{tab:ClassModels} Continued:} COVID-19 classification models.}\\
 \toprule
 \textbf{Ref.} & \textbf{Input data} & \textbf{Dataset size} & \textbf{Dataset diversity} & \textbf{Number of classes} & \textbf{Architecture} & \textbf{Bias and over-fitting prevention} & \textbf{Annotation dependency}\\
 \midrule
 \endhead
 Ref.~\cite{MeiLee} &  CT & 419 COVID-19; 486 Non-COVID & 18 centers & Binary & Inception and ResNet CNN & Data Augmentation; Transfer Learning & Lung segmentation\\
 \\
 Ref.~\cite{ArdakaniRajabzadeh} & CT & 108 COVID-19; 86 Non-COVID & One hospital & Binary & ResNet CNN & Transfer learning & Infection segmentation\\
 \\
 Ref.~\cite{Rahimzadeh:2020} & X-ray & 118 COVID-19; 6054 Non-COVID; 8851 Normal & Two public datasets & Multi-class & Xception and ResNet & Data Augmentation; Transfer Learning & Not required \\
 \
 Ref.~\cite{BaiWang} & CT & 521 COVID-19; 665 Non-COVID & COVID from 10 hospitals; Non-COVID from 3 hospitals & Binary & EfficientNet CNN & Data augmentation; Transfer learning & Lung segmentation\\
 \
 Ref.~\cite{CastiglioniIppolito} & X-ray & 250 COVID-19; 250 Non-COVID & 2 hospitals & Binary & ResNet CNN & Data augmentation; Transfer learning & Not required\\
 \\
 Ref.~\cite{LiQin} & CT & 4356 chest CT exams from 3322 patients & 6 hospitals & Multi-class & ResNet50 CNN & Data augmentation & Lung segmentation\\
 \\
 Ref.~\cite{AbbasAbdelsamea} & X-ray & 105 COVID-19; 11 Non-COVID; 80 Normal & Multi-center & Multi-class & CNN & Data augmentation; Transfer learning & Not required\\
 \\
 Ref.~\cite{WaheedGoyal} & X-ray &  403 COVID-19; 721 Normal & Multi-center & Binary & VGG CNN & Gan-based data augmentation; Transfer learning & Not required\\
 \\
 Ref.~\cite{AfsharHeidarian} & X-ray & 266 COVID-19; 5538 Non-COVID; 8066 Normal & Multi-center & Binary & Capsule network & Loss modification; Transfer learning & Not required\\
 \\
 Ref.~\cite{WangKang} & CT & 44 COVID-19; 55 Non-COVID & 3 hospitals & Binary & Inception CNN; Ensemble of classifiers & Transfer learning & Infection segmentation\\
 \\
 Ref.~\cite{WangWong} & X-ray & 266 COVID-19; 5538 Non-COVID; 8066 Normal & Multi-center & Multi-class & CNN & Data augmentation; Transfer learning & Not required\\
 \\
 Ref.~\cite{GozesFrid2} & CT & 106 COVID-19; 100 Normal & Multi-center & Binary & ResNet CNN & Data augmentation; Transfer learning & Lung segmentation\\
 \\
 Ref.~\cite{FarooqHafeez} & X-ray & 45 COVID-19; 1591 Non-COVID; 1023 Normal & Multi-center & Multi-class & ResNet CNN & Data augmentation; Transfer learning & Not required\\
 \\
 Ref.~\cite{NarinKaya} & X-ray & 50 COVID-19; 50 Normal & One public dataset & Binary & ResNet CNN & Transfer learning & Not required\\
 \\
 Ref.~\cite{OzturkTalo} & X-ray & 127 COVID-19; 500 Non-COVID; 500 Normal & Two public datasets & Multi-class; Binary & DarkNet (YOLO base model) & NA & Not required\\
 \\
 Ref.~\cite{KarimDohmen} & X-ray & 266 COVID-19; 5538 Non-COVID; 8066 Normal & Three public datasets & Multi-class & Ensemble CNN & Data augmentation & Not required\\
 \\
 Ref.~\cite{AmyarModzelewski} & CT &  449 COVID-19; 495 Non-COVID; 425 Normal & 3 hospitals & Multi-class & Encoder-decoder CNN & Multi-task learning & Not required\\
 \\
 Ref.~\cite{NarayanKumar} & CT & 127 COVID-19; 500 Non-COVID; 500 Normal & Two public datasets & Multi-class & Xception CNN & Transfer learning & Not required\\
 \\
 Ref.~\cite{IslamIslam} & X-ray & 1525 COVID-19; 1525 Non-COVID; 1525 Normal & Several public datasets & Multi-class & LSTM & NA & Not required\\
 \\
 Ref.~\cite{HemdanShouman} & X-ray & 25 COVID-19; 25 Non-COVID & Two public datasets & Binary & CNN & NA & Not required\\
 \\
 Ref.~\cite{ZhangXie} & X-ray & 599 COVID-19; 24622 Non-COVID; 18881 Normal & Multi-class & One-class anomaly detection & EfficientNet CNN & Data augmentation; Transfer learning & Not required\\
 \\
 Ref.~\cite{GozesFrid} & CT & 159 COVID-19; 90 Non-COVID & Multi-center & Binary & ResNet CNN & Data augmentation; Transfer learning & Lung segmentation\\
 \\
 Ref.~\cite{HuGao} & CT & 150 COVID-19; 150 Non-COVID; 150 Normal & Two hospitals & Multi-class & Multi-scale CNN & Loss modification; Data augmentation & Lung segmentation\\
 \\
 Ref.~\cite{YingZheng} & CT & 88 COVID-19; 101 Non-COVID; 86 Normal & hospitals of two provinces in China & Multi-class & CNN & NA & Lung segmentation\\
 \\
 Ref.~\cite{WangDeng} & CT &  540 COVID-19;  229 Normal & One hospital & Binary & CNN & Data augmentation & Lung segmentation\\
 \\
 Ref.~\cite{OhPark} & X-ray & 180 COVID-19; 131 Non-COVID;  191 Normal & Several public datasets & Multi-class & ResNet CNN & Transfer learning & Lung segmentation\\
 \\
 Ref.~\cite{SedikIliyasu} & X-ray; CT & 288 COVID-19; 288 Normal & NA & Binary & LSTM & Gan-based data augmentation & Not required\\
 \\
 Ref.~\cite{XuJiang} & CT & 110 COVID-19; 224 None-COVID, 175 Normal & 3 hospitals & Multi-class & CNN & Changing sampling probability & Infection segmentation\\
 \\
 Ref.~\cite{MohammedWang} & CT & 20 COVID-19; 282 Non-COVID & One public dataset & Multi-class & Bi-directional LSTM; CNN & Changing sampling probability; Data augmentation & Lung segmentation\\
 \\
 Ref.~\cite{YangJiang} & CT & 146 COVID-19; 149 Normal & One hospital & Binary & CNN & data augmentation & Lung segmentation\\
 \\
 Ref.~\cite{MengDong} & CT & 366 COVID-19 & 4 centers & Binary & 3D CNN with integrated clinical data & Loss modification; Data augmentation & Lung segmentation\\
 \\
 Ref.~\cite{HeidarianAfshar} & CT & 171 COVID-19; 60 Non-COVID; 76 Normal & One center & Binary & Capsule network & Loss modification & Lung segmentation\\
 \\
 Ref.~\cite{HeidarianAfshar2} & CT & 171 COVID-19; 60 Non-COVID; 76 Normal & One center & Binary & Capsule network & Loss modification & Lung segmentation\\
 \hline

 \end{longtable}
 \end{tiny}
 \end{center}
 \end{landscape}
% \restoregeometry
\twocolumn
%%%%%%%%%%%%%%%%%%%%%%%%%%%%%%%%%%%%%%

\vspace{.1in}
\noindent
\textbf{\textit{COVID-19 Classification without Annotation:}} Studies that do not include any segmentation as a pre-processing step essentially feed the developed model with raw images. As CXR images are single slices and simpler to process compared to CT scans, they are utilized without annotation in most of the studies. Reference~\cite{OzturkTalo} is an example of such studies, where raw CXR images are fed to a DL model for binary and multi-class COVID-19 classification. Narayan Das \textit{et al.}~\cite{NarayanKumar} and Islam \textit{et al.}~\cite{IslamIslam} also utilized CXR images without annotation for a three-way COVID-19 classification.  Although using CT scans, Reference~\cite{AmyarModzelewski} is independent from segmented inputs. It exploits annotation labels in the output layer to develop a multi-task training framework. In other words, both classification and segmentation are aimed at in this work.

\vspace{.1in}
\noindent
\textbf{\textit{Lung Segmentation for COVID-19 Classification:}} Lung segmentation is the first step in many COVID-19 classifications studies, as it eliminates unessential information. Gozes \textit{et al.}~\cite{GozesFrid}, for instance, used a pre-trained U-net model for this task. Since the segmentation model should be able to annotate lungs even in the presence of COVID-19 opacities, the U-net model is fine-tuned on a dataset of interstitial lung disease cases. More advanced lung segmentation models are also used in COVID-19 studies. Reference~\cite{HuGao}, for instance, has proposed a multi-window U-Net that incorporates several windows instead of the standard Hounsfield unit (HU) window. Furthermore, this study uses a sequential information attention module to integrate all CT slices.

\vspace{.1in}
\noindent
\textbf{\textit{Infection Segmentation for COVID-19 Classification:}} Beside lung segmentation, some COVID-19 classification studies rely on segmenting the pulmonary regions of infection. Xu \textit{et al.}~\cite{XuJiang}, for instance, have used a 3D CNN model trained on pulmonary tuberculosis for infection segmentation. Although this model is not trained on a COVID-19 dataset, it can still extract candidate patches. The annotation results are consequently used to form cubic patches around the regions of infection, which are then fed to the classification model. Based on the COVID-19 characteristics, such as GGO, Wang \textit{et al.}~\cite{WangKang} have manually delineated the CT scans to extract all the ROIs, from which 2-3 patches are randomly selected as the input to the CNN model for classification purposes.

%
%------------------------------------------------------------------------------------------------------
\subsubsection{COVID-19 Classification Types}
%------------------------------------------------------------------------------------------------------
Binary or multi-class COVID-19 classification refers to whether the problem is considered as COVID-19 versus all other possible categories as one class or all the classes are treated separately.  These two approaches are investigated in the following:

\vspace{.1in}
\noindent
\textbf{\textit{Binary COVID-19 Classification Problems:}} Reference~\cite{OzturkTalo} is an example of binary classification, where the goal is to distinguish between COVID-19 and non-COVID cases. Non-COVID cases include both normal and pneumonia patients. Reference~\cite{NarinKaya} explores three different COVID-19-related binary classification problems, in each of which COVID-19 is classified against a different class, including viral pneumonia, bacterial pneumonia, and normal. Obtained results show that COVID-19 is best distinguishable from bacterial pneumonia. Beside positive and negative COVID-19, patients can be classified based on other clinical outcomes. Meng \textit{et al.}~\cite{MengDong}, for instance, consider high and low-risk as the binary classification labels.

\vspace{.1in}
\noindent
\textbf{\textit{Multi-class COVID-19 Classification Problems:}} Reference~\cite{OzturkTalo}, beside considering a binary classification problem, tries to solve a multi-class classification consisting of three classes of COVID-19, pneumonia, and normal. The obtained accuracy, however, is lower than the binary scenario. The same categorization is followed in~\cite{IslamIslam}. Reference~\cite{AmyarModzelewski} also followed a three-way classification with the difference that all diseases other than COVID-19 are considered as the ``others" class to be classified against COVID-19 and normal subjects. COVID-19, pneumonia, and other diseases are considered as three separate classes in Reference~\cite{NarayanKumar}. Since Reference~\cite{XuJiang} have used annotated infection patches as inputs to a CNN model, it also considers a irrelevant-to-infection class to exclude incorrectly segmented areas.

It is worth mentioning that unlike binary and multi-class approaches, COVID-19 classification is considered as a one-class anomaly detection in Reference~\cite{ZhangXie}, where the model's output is the anomaly score of the input, along with a confidence score that determines the model's confidence in its prediction. Consequently, subjects with a high anomaly score or low confidence score are considered as positive COVID-19.

%------------------------------------------------------------------------------------------------------
\subsubsection{Imaging Modality used for COVID-19 Classification}
%------------------------------------------------------------------------------------------------------
CXR and CT are two common imaging modalities considered in the COVID-19 classification studies. These two modalities, however, require different processing strategies, as described below:

\vspace{.1in}
\noindent
(i) \textbf{\textit{COVID-19 Classification via CXR Images:}}  CXR images are 2D and as such processing techniques to incorporate the relation between images are not required. CXR images can be independent inputs to a DL model. References~\cite{OzturkTalo,NarayanKumar,IslamIslam} are  examples of using CXR images for classification tasks. Unlike most of the COVID-19 classification methods using CXR that incorporate the whole image at once, Oh \textit{et al.}~\cite{OhPark} extract several random patches from the input image and feed them individually to the DL model. The final decision is a majority voting over all the obtained outcomes.
\vspace{.1in}
\noindent
(ii) \textbf{\textit{COVID-19 Classification via CT Scans:}} Unlike CXR images, CT scans are 3D in the sense that each patient is associated with several 2D slices. As a result, analyzing CT scans require specific strategies, the first of which is a slice-level classification, where slices are treated independently with the goal of assigning labels to separate slices. Patient-level classification, on the other hand, tries to make the final decision using all the available slices.
\begin{itemize}
\item \textit{Slice-level Classification}: Reference~\cite{AmyarModzelewski}, as an example of a slice-level classification algorithm, uses separate slices as inputs to a DL model, where slices are gathered from three different data sources and pre-processed to have consistent size, resolution, and contrast. Reference~\cite{HuGao} has assigned patient-level labels to all the slices and leveraged a 2D CNN model. This strategy, however, can cause inconsistency when a slice without any visible manifestation is assigned with COVID-19 or pneumonia label. References~\cite{YangJiang,GozesFrid2} are other examples of slice-level classification models, where target slices are manually selected to train the CNN model. At the test time, however, these studies, average over all the probabilities to form the patient-level classification. Therefore, the underlying studies can be considered as cross-sections of slice-level and patient-level classification, bringing us to the discussion in the next part, i.e., patient-level classification.
\item \textit{Patient-level Classification}: Patient-level classification using CT scans requires a voting strategy to combine the slice-level outcomes. The voting mechanism is of particular importance as the whole CT volume cannot be typically processed at once. Different voting mechanisms have been developed in the literature including the following items:
\begin{itemize}
\item \textit{Volumetric Scoring:} In Reference~\cite{GozesFrid} 2D slices are first processed to form the slice-level outcomes. Summing over the activation maps of the detected positive slices, consequently, results in the aforementioned volumetric score. It is worth mentioning that only activations above a pre-defined threshold are considered in the summation. The obtained COVID-19 score can also be considered as the extent of the disease in a patient's lungs.
\item \textit{Pooling Operations:}
%
% Reference~\cite{YingZheng} is another example of
For patient-level classification, one approach~\cite{YingZheng} is to combine different models (e.g., parallel CNNs) in a parallel architecture. Results from individual slices can then be aggregated through pooling operations. Similarly, Li \textit{et al.}~\cite{LiQin} incorporate parallel CNNs, results of which are aggregated through a max pooling operation.
\item \textit{Whole CT Volume:} To leverage the information from all the CT scans and capture their relations, Wang \textit{et al.}~\cite{WangDeng} feed their developed CNN model with the whole CT volume, which is concatenated with the segmented lung mask. The same strategy of feeding the whole CT volume is also used in Reference~\cite{MengDong}.
\item \textit{Bayesian Merging:} A Noisy-or Bayesian function is adopted in Reference~\cite{XuJiang} to combine outcomes of several infection patches.
\item \textit{RNN-based Merging:} Using Recurrent Neural Networks (RNNs) is another strategy to combine the slice-level information and consider the spatial relations. This group of models are discussed in Section~\ref{sec:arch}.
\item \textit{Multi-stage Frameworks:}  Designing a multi-stage framework is also a common patient-level classification approach. Mei \textit{et al.}~\cite{MeiLee}, for instance, have designed a two stage workflow, where in the first stage abnormal slices are detected using a previous pre-trained pulmonary tuberculosis (PTB) detection model. Top ten candidate slices are then fed to another CNN, in stage two, to identify slices with positive COVID-19.  Final outcome is ultimately set as the average of slice-level prediction of a patient’s ten most abnormal candidates. The same multi-stage strategy is followed in Reference~\cite{HeidarianAfshar}, with the difference that the CNNs are replaced with capsule networks, as described in Section~\ref{sec:arch}.
\end{itemize}
\end{itemize}

%------------------------------------------------------------------------------------------------------
\subsubsection{DL Architectures for COVID-19 Classification}\label{sec:arch}
%------------------------------------------------------------------------------------------------------
Although different DL architectures are applicable to the task of image classification, in the COVID-19 scenario, discriminative models including CNNs, RNNs, and capsule networks are the most commonly used ones. These networks and how they are incorporated in COVID-19 classification studies are explained below.

\vspace{.05in}
\noindent
\textit{\textbf{CNN-based COVID-19 Classification Models:}} CNNs are stack of convolutional and pooling layers, often followed by fully connected ones. Since the trainable filters share weight across the whole image, these networks are computationally effective, and can extract local features from the input. CNNs have shown promising results in the field of image processing including COVID-19 classification. Although it is possible to design a CNN from scratch, most of the studies have built their models upon pre-existing successful CNN models as described below:
\begin{itemize}
\item \textit{Pre-existing CNN models}: Since the start of the outbreak, the following pre-existing CNN models for COVID-19 classification:
\begin{itemize}
\item \textit{Darknet-19 Model:} DarkCovidNet model proposed in Reference~\cite{OzturkTalo} is a modification of Darknet-19 model, which is the basis of YOLO object detection system. The proposed DarkCovidNet consists of $17$ convolutional layers, which are followed by pooling layers, and eventually one fully connected layer for the final classification.
\item \textit{Inception Model}, is another commonly used CNN model utilized in COVID-19 studies. Reference~\cite{NarayanKumar}, for instance, utilizes extreme version of the Inception model, referred to as Xception. Two other variations of Inception, referred to as InceptionV3 and Inception-ResNetV2, are exploited in Reference~\cite{NarinKaya}, along with three variations of the popular ResNet architecture, namely ResNet50, ResNet101, and ResNet152. Obtained results show superior performance for ResNet50 model.
\item \textit{ResNet Models:} ResNet50 is the basis of the model proposed in Reference~\cite{FarooqHafeez}, referred to as the COVID-ResNet. This model is trained in three stages, where in each stage the image size is increased gradually. The ResNet50 model of Reference~\cite{GozesFrid} is followed by a GradCam localization to verify the pathological areas focused through the training process. The resulting map can provide insights to the radiologist.
%
% \item \textit{VGG Models:} Beside the variations of ResNet~\cite{GozesFrid} and Inception, VGG models, such as VGG19~\cite{HemdanShouman}, are common choices.
%
%
\item \textit{EfficientNet}, utilizing compound coefficients to scale up CNNs, is another architecture used for COVID-19 classification in Reference~\cite{ZhangXie}.
\item \textit{Ensemble Models:} Besides adopting the pre-existing CNN architectures, it is also possible to develop ensemble frameworks to leverage the potentials of different CNN models.
%Such strategy is used in Reference~\cite{KarimDohmen}, where several models including VGG and ResNet are combined.
\end{itemize}
\item \textit{Self-designed CNN models}: Based on the identified requirements, some studies have designed their own specific CNN models for COVID-19 classification. A multi-scale CNN, for instance, is proposed in Reference~\cite{HuGao}, where intermediate CNN representations are aggregated through a global Max Pooling operation to make the final decision. The self-designed CNN model proposed by Wang \textit{et al.}~\cite{WangDeng} consists of three subsequent blocks, the first of which is a vanilla 3D CNN, followed by a residual block. The last part is a progressive classifier, containing convolutional and fully-connected layers. Beside focusing on designing  layers of a CNN, another strategy is to feed the model with information other than the raw image. Such strategy is leveraged in Reference~\cite{XuJiang}, where the distance between the center of infection and pleura is concatenated with a fully-connected layer. This distance can contribute to a more accurate classification, as COVID-19 infection has a pleural distribution, partly distinguishing it from other diseases. Meng \textit{et al.}~\cite{MengDong} utilized patient's clinical factors, such as gender, age, and chronic disease history, as the additional information to be concatenated with the CNN's fully connected layer. More heterogeneous factors, including travel and exposure history and symptomatology, are incorporated in the model designed by Mei \textit{et al.}~\cite{MeiLee}.
\end{itemize}

\vspace{.05in}
\noindent
\textit{\textbf{RNN-based COVID-19 Classification Models:}} RNNs are especially useful in medical imaging when the goal is to process the whole volume or analyze follow-up studies. Since RNNs are subject to the problem of vanishing gradients, LSTM networks are commonly used as an effective alternative. The vanilla LSTM is not designed for extracting local features from images and as such this network is often combined with a CNN, to make use of its weight sharing advantages. Such a model is utilized in Reference~\cite{IslamIslam} for COVID-19 classification, resulting in a CNN-LSTM design. In the underlying study, $12$ convolutional layers are first incorporated to extract features from CXR images. The output of the CNN is then fed to an LSTM, the result of which determines the probability of COVID-19, pneumonia and normal classes. While a conventional LSTM considers only forward relations, bi-directional LSTMs additionally take the backward relations into account. Such models are incorporated in Reference~\cite{MohammedWang} for COVID-19 classification.

\vspace{.05in}
\noindent
\textit{\textbf{CapsNet-based COVID-19 Classification Models:}} CapsNets are relatively new deep learning architectures, proposed to solve the incapability of CNNs to recognize spatial information. Each capsule in a CapsNet, consist of several neurons to represent an object instantiation parameters, as well as its existence probability. The main feature of the CapsNet is its routing by agreement process, through which capsules in a lower layer predict the outcome of capsules in the next layer. The parent capsules take these predictions into account, based on the similarity (agreement) between the prediction and actual outcome. Using the routing by agreement, CapsNet is capable of recognizing spatial relations between image instances, and therefore handle much smaller datasets, compared to CNNs. Reference~\cite{AfsharHeidarian} has recently exploited CapsNets for the problem of COVID-19 classification using CXR, showing improvements over the CNN counterparts. The proposed architecture, referred to as COVID-CAPS, consists of several convolutional, pooling, and capsule layers, the output of which determines the probability of positive COVID-19.

% A similar architecture is utilized in Reference~\cite{HeidarianAfshar2}, with the difference that CT scans are used, and as such slice-level final capsules are combined to form a patient-level decision.
%ARASH: Add COVID-FACT

%OOOOOOOOOOOOOOOOOOOOOOOOOOOOOOOOOOOOOOOOOOOOOOOOOOOOOOO
\section{Challenges, Open Problems, and Opportunities} \label{sec:COO}
%OOOOOOOOOOOOOOOOOOOOOOOOOOOOOOOOOOOOOOOOOOOOOOOOOOOOOOO
%%%%%%%%%%%%%%%%%%%%%%%%%%%%%%%%%%%%%%%
\begin{figure*}[t!]
\centering
\includegraphics[width=0.7\textwidth]{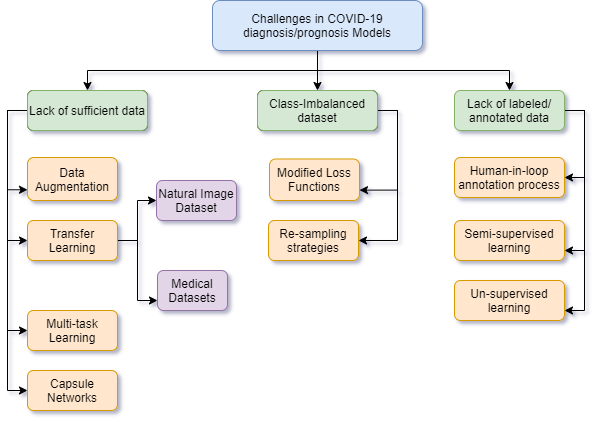}
\caption{Challenges and possible solutions in developing COVID-19 diagnosis/prognosis Models.}
\label{fig:tax_challenge}
\end{figure*}
%%%%%%%%%%%%%%%%%%%%%%%%%%%%%%%%%%%%%%%
In this section, first, we focus on limitations and challenges of developing COVID-19 diagnosis/prognosis models as shown in Fig.~\ref{fig:tax_challenge}. Then, we discuss open problems and potential opportunities for SP research by highlighting problems and challenges of developing SP/DL models for COVID-19 management.

%===================================================
\subsection{Challenges in Developing COVID-19 Diagnosis/Prognosis Models}
%===================================================
The ultimate goal of developing COVID-19 diagnosis/prognosis models is to be used in clinical applications and reduce the healthcare system's workload during pandemic conditions. Some  models proposed for diagnosis and prognosis of COVID-19 have shown successful results in real applications and enhanced the performance of junior radiologists to senior-level~\cite{ZhangXiaohong}. However, some common issues such as the risk of bias and over-fitting may cause poor generalization of such models.  The leading causes of these issues are: (i) Lack of sufficient data; (ii) Lack of labeled/annotated data, and; (iii) Imbalanced dataset. These three categories are described below together with solutions developed in COVID-19 studies to overcome them:
\subsubsection{Lack of Sufficient Data:} There is no doubt that preparing a high-quality dataset is the most critical part of developing a data-driven model. Collection of sufficient data for training robust COVID-19 models is challenging because: (i) COVID-19 is a new arising disease; (ii) Restrictions imposed to preserve patients' privacy, and; (iii) Health centers' strict data sharing protocols. On the other hand, despite the robustness of CNNs in hierarchically extracting high-value features from images, they cannot recognize the spatial relationships between those features. Due to various shapes and complex appearances of COVID-19 lesions, a large number of chest medical images are required to avoid over-flitting and ensure the model generalization. Data augmentation, transfer learning, multi-task learning, and the use of Capsules networks are some solutions to tackle these problems.

\vspace{.05in}
\noindent
\textit{Data Augmentation}, compensates for the lack of large training dataset, by generating several variations of the original samples~\cite{GozesFrid}. Random cropping, zooming, and flipping, for instance, are applied to the samples in Reference~\cite{ZhangXie}. Flipping, random rotation, and lighting are also adopted in Reference~\cite{FarooqHafeez} to further enlarge the training set. Random Gaussian noises is another data augmentation technique that has been used in Reference~\cite{WangLiu}. Other than applying different transformations to the dataset, Generative Adversarial Networks (GANs) can be used to generate new instances~\cite{SedikIliyasu}.
GANs produce fake images and forces the algorithm to discriminate them from the original ones, which makes the model robust on unseen images. Such strategy is utilized in Reference~\cite{SedikIliyasu}, where a convolutional GAN is leveraged for COVID-19 data augmentation. Similarly, a conditional GAN, referred to as COVIDGAN, is proposed in Reference~\cite{WaheedGoyal}, for CXR data augmentation.

\vspace{.05in}
\noindent
\textit{Transfer Learning,} refers to pre-training a model using an external dataset with the goal of encouraging the model to learn meaningful filters. The model is then fine-tuned on the main dataset, which might be a small one for an independent training. Transfer learning has shown promising results especially in field of medical imaging, where large datasets are scarce. While most of the existing studies utilize natural image datasets for pre-training, it is also possible to leverage similar medical samples as described below:
\begin{itemize}
\item \textit{Natural Image Dataset:} Reference~\cite{NarayanKumar} utilized transfer learning to fine-tune a pre-trained Xception model using a COVID-19 dataset. Transfer learning is also explored in Reference~\cite{NarinKaya} to pre-train five well-known CNN models, namely ResNet50, ResNet101, ResNet152, InceptionV3 and Inception-ResNetV2. ImageNet is the common choice for pre-training the CNN models~\cite{ZhangXie,FarooqHafeez,GozesFrid}. Some COVID-19 segmentation models incorporated an encoder pre-trained on ImageNet in their segmentation models to achieve more accurate results~\cite{QiuLiu}.
\item \textit{Medical Datasets:} Although pre-training with natural image datasets is very common in COVID-19 classification, it is also possible to leverage similar medical datasets, having the advantage of providing more useful filters and features. Such strategy is recently adopted by Afshar \textit{et al.}~\cite{AfsharHeidarian}, where the proposed CapsNet is pre-trained on a CXR dataset collected for a completely different task. In the fine-tuning phase, all the convolutional layers are kept fixed and only the capsule layers are re-trained on the COVID-19 dataset. Reference~\cite{MaWang} trained their COVID-19 segmentation network on a dataset containing $80$\% of lung cancer CT images and $20$\% of COVID-19 CT samples. However, the model failed to segment the COVID-19 regions of infection in test set due to the significant appearance differences between lung cancer tumors and COVID-19 lesions.
\end{itemize}

\vspace{.05in}
\noindent
\textit{Multi-task Learning} is a popular strategy to leverage the information available in several related tasks and make use of small datasets associated with different end goals. Multi-task learning is shown to be effective in reducing over-fitting and can be further divided into two categories, i.e., (i) Hard parameter sharing, and; (ii) Soft parameter sharing. In the former category, different tasks explicitly share several layers. In the latter, however, separate models are trained for separate tasks and the parameters are encouraged to take close values. With this in mind, Amyar \textit{et al.}~\cite{AmyarModzelewski}  proposed a DL model to perform COVID-19 classification, segmentation and reconstruction at the same time, using the hard parameter sharing strategy. The model begins with an encoder to encode all CT scan into a latent space for subsequent analysis. For the segmentation and reconstruction tasks, the latent space is decoded into the original feature space. In the classification scenario, however, the latent space goes through a MLP for the final three-way classification. It is also worth mentioning that the encoder-decoder architecture follows the well-known U-Net design. While Mean Squared Error (MSE) is utilized for the reconstruction part, dice score and cross-entropy losses are adopted for segmentation and classification, respectively. The final loss is the sum over all the three losses.

\vspace{.05in}
\noindent
\textit{Capsule Networks}, which are less data-demanding in comparison to CNNs and can be trained using smaller datasets. Incorporation of Capsule Networks for COVID-19 diagnosis models and their superiority when having a limited dataset at hand has been discussed in~\cite{AfsharHeidarian}. Capsule networks can be adopted instead of CNNs in medical segmentation networks. Capsule network-based segmentation network can potentially outperform its CNN-based counterparts. Due to data limitations for COVID-19 lesion segmentation, there is a potential for further investigation of replacing CNNs with Capsule Networks in segmentation models.

%--------------------------------------------------------------------------------------------------------
\subsubsection{Class-Imbalanced Dataset}
%--------------------------------------------------------------------------------------------------------
This problem occurs in situations where samples associated with one class outnumber those of the other class, which is a common problem in most real-world problems, including COVID-19 diagnosis/prognosis. In training a model for diagnosis of COVID-19 from normal/CAP cases, we usually have a fewer number of COVID-19 instances and a larger number of other classes. It is the same situation in segmentation models, where the pixels labeled as COVID-19 lesions are in the minority compared to the background pixels. In such scenarios, the model can be biased toward the majority class. To tackle this issue, one can consider: (i) \textit{Modified loss functions}, or; (ii) \textit{Re-sampling techniques} as possible solutions, as discussed below:

\vspace{.05in}
\noindent
\textit{Modified Loss Functions}, which improve the model performance by assigning more penalty to the mis-classified instances/pixels of the minority class. Weighted binary cross-entropy is a commonly used loss function in class-imbalanced classification/segmentation models~\cite{RajamaniSiebert}. Focal Tversky loss is a re-weighted loss functions that optimizes the coverage of predicted and ground-truth masks by assigning more weights to the target pixels. Some COVID-19 segmentation studies adopt a combination of modified loss functions to improve the model performance in both image-level and small ROIs~\cite{FanZhou}. Reference~\cite{HuGao,MengDong} developed a COVID-19 diagnosis model under supervision of a focal loss which assigns smaller weight to easy examples, and thus they contribute less to the loss function.

\vspace{.05in}
\noindent
\textit{Re-sampling Strategies}, that handle the class imbalance problem by either over-sampling the minority class instances or under-sampling from the majority class.  Xi \textit{et al.}~\cite{OuyangHuo} adopted an over-sampling strategy in their COVID-19 diagnosis model to adjust the samples of different classes in each mini-batch.
Li \textit{et al.}~\cite{LiWei} introduced a new off-line sampling  strategies that ranks the non-COVID-19 samples based on their diversity and difficulty. The most informative samples are then fed into the classification model.  Their approach could significantly decrease the training time while achieving comparable results.

%--------------------------------------------------------------------------------------------------------
\subsubsection{Lack of Labeled/Annotated Data}
%--------------------------------------------------------------------------------------------------------
One of the most challenging problems when developing medical segmentation networks is the lack of pixel-level labeled images. Pixel-level labeling of medical images by experienced radiologists is time-consuming. When it comes to COVID-19 infections segmentation, since the regions of infection are blurry with hardly-distinguishable boundaries from healthy lung tissues, the experts' annotation may not be consistent, making it necessary to work with a team of radiologists. Manual annotation of one COVID-19 CT volume takes $1$ to $5$ hours~\cite{ShanGao}. Some helpful solutions to overcome this problem are as follows:

\vspace{.05in}
\noindent
\textit{Human-in-loop Annotation Process}, which is a human-machine collaboration approach to ease and accelerate the annotation process. Shan \textit{et al.}.~\cite{ShanGao} proposed an efficient human-in-the-loop system by the collaboration of radiologists and the DL-segmentation model, which could dramatically reduce the annotation time.

\vspace{.05in}
\noindent
\textit{Semi-supervised Learning}, where a few annotated samples together with a large number of non-annotated CT images are fed to the network to increase the model accuracy. Combining 3D segmentation with GANs in a semi-supervised fashion can also be used to segment COVID-19 lesions. GASNet proposed in~\cite{XuCao} is a 3D segmentation framework containing a segmentation network with embedded GANs and a discriminator that uses a semi-supervised approach to segment the COVID-19 lesions. The experimental results on three external public datasets showed that their model's performance trained on a few labeled CT volumes was comparable with fully-supervised segmentation networks trained on a large labeled dataset.

\vspace{.05in}
\noindent
\textit{Un-supervised Learning}, where the model distinguishes out of distribution data in a dataset without any pre-existing label. Li \textit{et al.}~\cite{LiWei} used a new paradigm of un-supervised learning, refereed to as self-learning, to exploit helpful information from unlabeled data in their COVID-19 classification model.

%=====================================================
\subsection{Open Problems}%=====================================================
In this section, we focus on open problems and potential opportunities for SP research by highlighting problems and challenges of developing SP/DL models for COVID-19 management.

\vspace{.05in}
\noindent
$\bullet$ COVID-19 patients suffer from dyspnea as such there are inevitable motion artifacts in the acquired images. This is in contrast to most of other medical images, where motion artifact is rarely present. The artifacts in the COVID-19 images sometimes overlap with the main areas of infection, making the diagnosis/prognosis challenging even for experienced radiologists. To eliminate the effect of artifacts, most of the studies simply remove the noisy data from the dataset. This, however, reduces the generalizability and applicability of the model in clinical practice. An alternative solution is to approach advanced artifact reduction techniques, among which adaptive techniques are of higher capability as they can adjust and track the signal under noisy conditions.

%%%%%%%%%%%%%%%%%%%%%%%%%%%%%%%%%%%%%%
\begin{figure}
\centering
\includegraphics[scale=0.5]{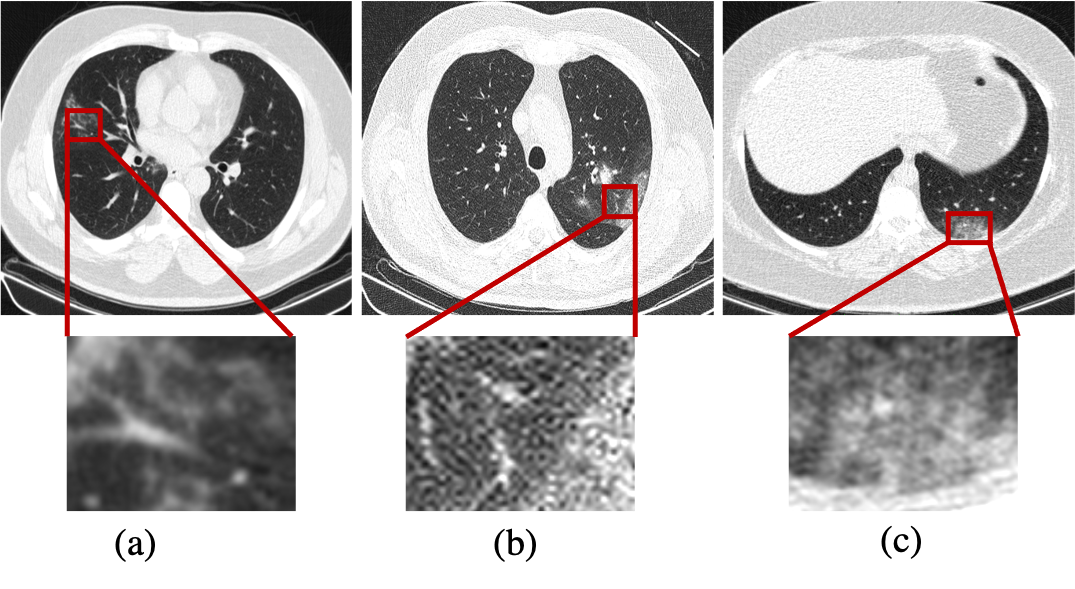}
\caption{Acquired CT images with three different dose levels for three COVID-19 patients. (a) Standard-dose, (b) Low-dose, (c) Ultra Low-dose.}\label{fig:dose}
\vspace{-.2in}
\end{figure}
%%%%%%%%%%%%%%%%%%%%%%%%%%%%%%%%%%%%%%
%%%%%%%%%%%%%%%%%%%%%%%%%%%%%%%%%%%%%%
\begin{figure*}[t!]
\centering
\includegraphics[scale=0.7]{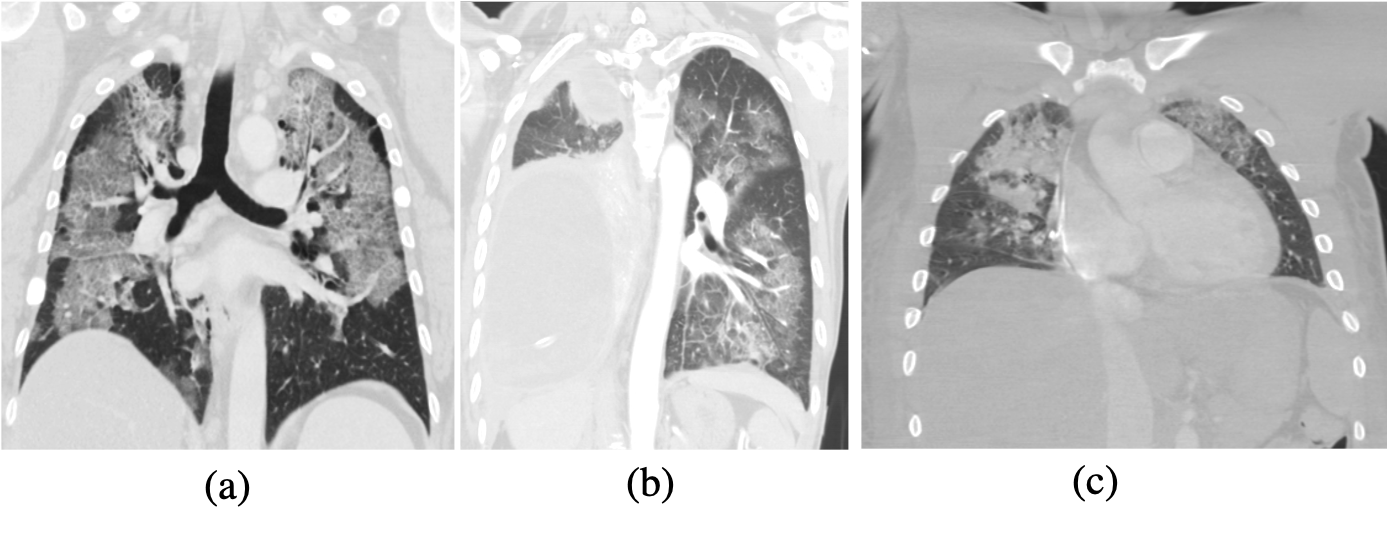}
\caption{Acquired images for three patients suspected to COVID-19. These patients have pre-existing conditions interfering with the diagnosis of COVID-19: (a) 47 year old male with fatty embolism and pulmonary edema, (b) 51 year old female with history of right lung cancer and right lower lobectomy, (c) 27 year old male with gunshot injury in the left hemithorax with pulmonary contusion and left hemothorax and pneumothorax.}
\label{fig:heart}
\vspace{-.2in}
\end{figure*}
%%%%%%%%%%%%%%%%%%%%%%%%%%%%%%%%%%%%%%
\vspace{.05in}
\noindent
$\bullet$ COVID-19 involves a large volume of the lung and is sparsely distributed around the lung volume. This is, in particular, in contrast to medical images in which the region of interest is located in a specific location of the organ. Analyzing and extracting patterns from the COVID-19 images require sparse filtering techniques within the signal processing domain.

\vspace{.05in}
\noindent
$\bullet$ As COVID-19 infection is distributed in the whole lung volume, the relation between the image slices is of high diagnostic and prognostic importance, calling for specific 3D filtering and pattern recognition approaches.

\vspace{.05in}
\noindent
$\bullet$ A key issue with chest CT scan is exposing patients to harmful radiation. In this regard, using low-dose or ultra low-dose scanning is of high interest. In a recent study by Tabatabaei \textit{et al.}~\cite{TabatabaeiTalari}, it is shown that obtained low-dose CT scans have high agreement with standard-dose ones, in terms of typical findings of COVID-19. More importantly, low-dose examinations are associated with less cancer risk, especially in young women. Fig.~\ref{fig:dose} shows the obtained CT scans for three different patients at three different dose levels, i.e., standard, low, and ultra low. According to this figure, although low and ultra low-dose images have more visible artifacts, they can still reveal the presence of COVID-19 infection. The artifacts, however, can hamper the effective training of the model. Furthermore, collecting a dataset of low-dose scans may not resolve this issue, as besides dose, other factors such as the patient's weight can influence the quality of the image, leading to a wide variety of possible artifacts.
This calls for SP/DL models that can cope with  images at different resolutions while providing the same level of diagnosis/prognosis performance.
%Permission statement.

\vspace{.05in}
\noindent
$\bullet$ COVID-19 is relatively new, and as such, large datasets are not easily accessible. Therefore, the developed SP/DL models should be capable of handling small datasets and yet capturing informative features.

\vspace{.05in}
\noindent
$\bullet$ To encourage physicians and health professionals to confidently utilize DL models, it is important to provide explanation and interpretations on the internal behaviour of the DL models and the achieved results and therefore eliminate the ``black-box perception''. Regarding the black-box nature of the deep learning models, communicating explainable outcomes to the physicians is essential for clinical adoption of implemented DL models. Several explainability techniques are leveraged in COVID-19 studies, the simplest of which is to verify the outcomes with a radiologist. This approach is, however, time-consuming and burdensome. Techniques, providing heat-maps of the most important regions of the input image, are also popular within the COVID-19 studies. One of the commonly used heat-map techniques is Class Activation Mapping (CAM), utilized in Reference~\cite{HuGao} at different feature levels. Gradient-weighted Class Activation Mapping (Grad-CAM), visually depicting the deep model's decision, is also a CAM approach with the advantage of not requiring re-training. Grad-CAM outcome shows how the developed model pays more attention to the regions of infection of the chest radiographs in~\cite{OzturkTalo,IslamIslam}. Saliency map has also shown interpretable outcomes within the COVID-19 studies~\cite{HuGao}.

Despite the advances in improving the explainability of the models, there are still examples for which the model fails to provide a clear explanation. Furthermore, heat-maps do not provide enough explanation of the unique features they used to distinguish between COVID-19 and CAP cases.

\vspace{.05in}
\noindent
$\bullet$ Due to the policy of protecting people's privacy and also immediate quarantine of mild cases without further examinations, scans with non-severe symptoms are missing from most of the public datasets, and models are mostly developed  based on patients with severe lung lesions who are at late/advanced stages of the disease. The models, therefore, are biased towards severe cases and cannot be easily generalized.

\vspace{.05in}
\noindent
$\bullet$ Evaluating the developed SP/DL models' performance in an unseen domain, results in a decrease in the sensitivity of COVID-19 diagnosis. Most of the developed models, however, incorporate data coming from a single hospital, without a cross-center validation. In other words, the impact of equipment differences are not fully considered yet, and data from different sources are required to verify the generalizability of the models.

\vspace{.05in}
\noindent
$\bullet$ One limitation associated with many COVID-19 studies is that they try to distinguish COVID-19 cases from normal ones or categorize normal and non-COVID pneumonia cases as one class. Studies who consider a separate CAP class also report a relatively poor performance in distinguishing the COVID-19 and CAP classes. This calls for developing models with stronger backbone architectures and higher capacities. Furthermore, pneumonia incidence samples are older compared to the COVID-19 ones and images from pneumonia patients with COVID-19 symptoms are not included in the datasets.

\vspace{.05in}
\noindent
$\bullet$ Although hybrid models, combining images and other relevant clinical information, can play an important role in COVID-19 analysis, few datasets are accompanied with demographic and clinical risk factors.

\vspace{.05in}
\noindent
$\bullet$ One important challenge associated with COVID-19 analysis is the disease manifestation in patients with complications other than COVID-19. Several diseases can impact the lung tissue and interfere or change the appearance of COVID-19. Interstitial lung diseases, pleural or cardiac diseases may have imaging manifestations that may mask superadded COVID-19, and make it challenging for the interpreting radiologist.  As shown in Fig.~\ref{fig:heart}, it is not clear if the abnormalities are related to COVID-19. This calls for developing more advanced SP solutions and unique features to facilitate COVID-19 identification.

%OOOOOOOOOOOOOOOOOOOOOOOOOOOOOOOOOOOOOOOOOOOOOOOOOOOOOOOOO
\vspace{-.1in}
\section{Conclusion} \label{sec:Conc}
%OOOOOOOOOOOOOOOOOOOOOOOOOOOOOOOOOOOOOOOOOOOOOOOOOOOOOOOOO
Medical imaging plays an important role in the diagnosis and management of COVID-19 infection. Signal Processing (SP) methods coupled with Deep Learning (DL) models can help to develop robust autonomous solutions for diagnosis/prognosis of COVID-19 based on chest images. In this article, an integrated sketch is presented for designing and developing intelligent models for the COVID-19 infection diagnosis/prognosis. Advanced SP methodologies and DL models for diagnosis and prognosis of COVID-19 are presented, taking into consideration major challenges and opportunities. This article  provides the SP community with a comprehensive introduction to various solutions to COVID-19 Radiomics. In addition, the article provides the required radiological background, available resources, and challenges/opportunities for extensive future SP research in this multidisciplinary domain to serve our diligent role in combating COVID-19 pandemic and possible  future similar ones.

%OOOOOOOOOOOOOOOOOOOOOOOOOOOOOOOOOOOOOOOOOOOOOOOOOOOOOOOOO
\section*{Acknowledgement}
%OOOOOOOOOOOOOOOOOOOOOOOOOOOOOOOOOOOOOOOOOOOOOOOOOOOOOOOOO
This project was partially supported by the Department of National Defence's Innovation for Defence Excellence and Security (IDEaS) program, Canada. 

We would like to thank the consulting committee and EiC of IEEE SPM for their two-round reviews and encouraging comments.

%OOOOOOOOOOOOOOOOOOOOOOOOOOOOOOOOOOOOOOOOOOOOOOOOOOOOOOOOO
%\bibliographystyle{IEEEbib}
\vspace{-.2in}
\bibliographystyle{plain}
%\bibliography{A_Ref} %SPM (Radiology2)

\end{document}